\pdfoutput=1

\documentclass{article}

\usepackage{bm,amssymb}
\usepackage{graphicx}
\usepackage{caption}
\usepackage{subcaption}

\usepackage{amsmath}
\usepackage{fullpage}
\usepackage{booktabs}

\usepackage{pdfpages}

\begin{document}


# Readout and control of a single nuclear spin with a meta-stable electron spin ancilla

Sang-Yun Lee[1+], Matthias Widmann[1], Torsten Rendler[1], Marcus Doherty[2]*, Thomas M. Babinec[3,5], Sen Yang[1], Moritz Eyer[1], Petr Siyushev[1], Birgit J. M. Haussmann[3], Marko Loncar[3], Zoltán Bodrog[4], Adam Gali[4], Neil Manson[2], Helmut Fedder[1]*, Jörg Wrachtrup[1]

1. 3. Physikalisches Institut and Research Center SCOPE, University Stuttgart, Pfaffenwaldring 57, 70569 Stuttgart, Germany
2. Laser Physics Centre, Research School of Physics and Engineering, Australian National University, Canberra, A. C. T., 0200, Australia
3. School of Engineering and Applied Science, Harvard University, Cambridge, Massachusetts 02138, USA
4. Department of Atomic Physics, Budapest University of Technology and Economics, Budafoki út 8, H-1111, Budapest, Hungary
5. Department of Applied Physics, Stanford University, 348 Via Pueblo Mall, Stanford, CA 9430 (current address)

+These authors contributed equally

*e-mail: helmut.fedder@gmail.com

*e-mail: marcus.doherty@anu.edu.au

**Electron and nuclear spins associated with point defects in insulators are promising systems for solid state quantum technology[1–3]. While the electron spin usually is used for readout and addressing, nuclear spins are exquisite quantum bits[4,5] and memory systems[3,6]. With these systems single-shot readout of nearby nuclear spins[5] as well as entanglement [4,7,8] aided by the electron spin has been shown. While the electron spin in this example is essential for readout it usually limits nuclear spin coherence[9]. This has set of the quest for defects with *spin-free* ground states[8,10]. Here, we isolate a hitherto unidentified defect in diamond and use it at room temperature to demonstrate optical spin polarization and readout with exceptionally high contrast (up to 45%), coherent manipulation of an individual excited triplet state spin, and coherent nuclear spin manipulation using the triplet electron spin as a meta-stable ancilla. By this we demonstrate nuclear magnetic resonance and Rabi oscillations of the uncoupled nuclear spin in the spin-free electronic ground state. Our study demonstrates that nuclei coupled to single metastable electron spins are useful quantum systems with long memory times despite electronic relaxation processes.**

Coupled electron and nuclear spins in solids are promising candidates for quantum memories and registers and present a particular class in the emerging field of hybrid quantum systems[7,11–14], in which different types of qubits perform distinct functions according to their advantageous properties. In this particular case, the nuclear spin, which is weakly coupled to the environment, serves as a long-lived

quantum memory, whereas the electron spin, which has a short coherence time, but interacts strongly with external fields, serves as a readout gate[4,7,9]. In this architecture, the permanent presence of the electron spin is a source of decoherence to the nuclear spin[8]. To avoid such electron spin induced decoherence, the electron that carries the spin should be physically removed when it is not needed and returned to the nuclear spin only when the initialization or readout gate is applied. One approach to address this problem is to frequently remove the electron, e.g. by photo-ionization with a strong laser, resulting in a motional narrowing type of decoupling[9]. Another approach – that is followed in this work – is to choose an electronic system with a spin-free ground state and a meta-stable excited spin state that can be optically pumped and activated prior to the application of the readout gate[8]. Such a system implemented with a single solid state defect would enable a universal nuclear spin gate. In the following we present a suitable defect in diamond and demonstrate coherent control of a single nuclear spin in the spin-free electronic ground state. Notably, besides the NV center, this is the second known single spin defect in the solid state, whose coherent spin motion is detectable at room temperature.

The defect[1] is observed in ultra-pure single crystalline HPHT diamond after fabrication of vertical nano-wires (see Fig. 1a,b), whose main purpose is to provide high photon collection efficiency[15]. The wires are etched into the sample surface by reactive ion etching (see SI). Fig. 1c shows a confocal fluorescence microscope image of the sample. To identify candidates for meta-stable triplet spin defects, we observe the fluorescence intensity of bright spots on the sample surface with and without a strong applied magnetic field from a permanent magnet (≈0.5 T) (see Fig. 1d). An arbitrary oriented magnetic field causes spin mixing and thereby modifies the steady-state fluorescence[16]. We observe two types of emitters with decreasing and increasing fluorescence response, corresponding to NV- centers and the new ST1 defects, respectively. The increase of the steady-state fluorescence is due to the mixing of meta-stable triplet states as will be shown later.

To characterize the defect further, we record the photon auto-correlation and observe a dip well below 0.5 at zero delay, which confirms that we indeed observe single defects[17] (see Fig. 1g). At higher excitation intensity, the photon auto-correlation shows pronounced 'bunching-shoulders', which supports the idea of the existence of a long-lived shelving state. To characterize the defect fluorescence, we present fluorescence emission spectra at room-temperature and at 10K (see Fig. 1e). In both cases a strong resonant emission line is observed at 550(1) nm, with a broad phonon side band extending to the near infrared. Notably, this resonant transition does not correspond to a known diamond defect line[18]. To characterize the strength of the optical transition, we present the excited state lifetime that we have determined in a room temperature experiment by pulsed optical excitation (see Fig. 1f). An excited state lifetime of 9.5(2) ns is observed.

Next, we search for a signature of an electron spin resonance in the optical fluorescence. Fig. 2a shows the fluorescence intensity that is observed while sweeping the microwaves from 200 to 1400 MHz. Three distinct lines marking an increase in fluorescence intensity upon spin resonance are observed at frequencies 2E = 278(1) MHz, D-E=996(1) MHz, D+E=1274(1) MHz. This optically detected magnetic resonance (ODMR) spectrum indicates a triplet electron spin with zero field splitting parameters[19]

---

[1] hereafter termed ST1-defect following the convention that is commonly applied in the EPR community

D=1135(1) MHz and E=139(1) MHz. The observed large E parameter could originate from the presence of a large local strain or from an intrinsic low symmetry of the defect. We have compared the spin resonance spectra of ~20 individual defects (see SI) and observed variations of the E parameter smaller than 5 MHz, indicating that the E originates from an intrinsic low symmetry of the defect. From ODMR spectra with a variable axial magnetic field applied (Fig. 2b), we determine an electronic $g$-factor $g_e = 2.0(1)$, close to the $g$-factor of the free electron spin. The absence of an intrinsic hyperfine splitting indicates that the defect's constituent atoms possess no nuclear spin (see SI for a detailed discussion of the possible constituents and structure of the defect). The aligned axial magnetic field has an angle of 35(1)° w.r.t the [111] crystal direction, indicating that the defect's main symmetry axis is oriented along the [110] crystal direction (see inset in Fig. 2b).

All findings reported up to here are compatible with two possible electronic configurations: the triplet state could either be the long lived shelving state or the ground state. To show that the triplet state is indeed the long lived shelving state, we measure the decay of the shelving state population. If the shelving state was a singlet, a single exponential decay would be expected. By contrast, in the present case, three distinct exponential decay terms are observed (Fig. 3b) and their amplitudes are sensitive to microwave fields resonant with the triplet ESR transitions. In each decay curve shown, the dominant decay terms are controlled by switching the spin population with microwave π-pulses resonant with the three ESR transitions. This indicates that the shelving state is indeed the triplet state. The result can be used to assign the decay constants to the spin quantum numbers (Fig. 3c). A complete model of the optical cycle can be established by determining also the triplet population rate by an analogous pulsed experiment (see SI). Fig. 3d summarizes all results related to the triplet kinetics. The essential features are: a rapid (~1/70 ns) inter-system-crossing rate from the singlet excited state to the triplet state and substantial differences among the triplet lifetimes, $\tau_{|0\rangle}$≈2500 ns, $\tau_{|-1\rangle}$≈1000 ns, $\tau_{|+1\rangle}$≈200 ns. These are responsible for the positive and exceptionally high (45 %) optical spin signal contrast, and increasing steady-state fluorescence response to a randomly aligned static magnetic field as observed in Fig. 1d because mixing the triplet sublevels results in faster decay to the ground state. It follows that optical pumping prepares the system in a spin polarized state of the meta-stable triplet state with predominant population of the slowly decaying triplet sublevel, |0⟩. We now apply coherent spin manipulation to demonstrate Rabi oscillations between the triplet spin states (Fig. 4). Because the spin population of the faster decaying states |+⟩ or |-⟩ is increased by a resonant ESR pulse, any coherent spin manipulation results in the enhancement of the fluorescence intensity as shown in Fig.4b. The oscillations are described by two exponential decay functions, one of which is modulated by a cosine. The decay times and oscillation amplitudes correspond to the average of the lifetimes and the difference of the initial populations of the two involved spin states, respectively, and are in excellent agreement with the rates measured independently above (Fig. 3d). Note that the decay times are fully accounted for by the triplet lifetimes, which implies that spin relaxation is negligible. The triplet lifetimes are long enough to coherently address a coupled nuclear spin. We have compared the Zeeman sweeps of about 20 defects and observe that two of them show hyperfine splittings due to proximal $I = 1/2$ carbon nuclear spins of several tens of MHz see (Fig. 5a).

We now show the application of the metastable electron spin as a universal single nuclear spin gate and demonstrate coherent manipulation and readout of the nuclear spin in the spin-free electronic ground state. To achieve this, we exploit the metastable electronic triplet spin ancilla to transfer spin-polarization onto and off the nuclear spin. To this end we exploit the optical nuclear polarization (ONP) induced by a level-anti-crossing (LAC) between the |-1,+1/2⟩ and |0,-1/2⟩ hyperfine levels at about 42 mT to selectively mix electronic and nuclear spin states[2020] (see SI). Under these conditions, spin-polarization is transferred to the nuclear spin when the electron enters the meta-stable triplet level resulting in optical polarization of the nuclear spin[21]. After the optical pump pulse is turned off, the electron decays to the singlet ground state within a time given by the triplet lifetime. This decay eliminates the triplet electron spin and leaves the nuclear spin in its polarized state. Likewise, when the system is optically pumped from the singlet ground state to the excited triplet state, the nuclear spin state is preserved and the hyperfine levels are populated correspondingly. At the LAC, the different lifetimes of the hyperfine levels result in different fluorescent intensity of the nuclear spin projections. This provides a mechanism to readout the nuclear spin (see Fig. 5b and Methods). Fig. 5c shows the nuclear spin signal contrast as the energy levels are tuned into the LAC. Nuclear spin polarization is maximal at the LAC and vanishes when the detuning from the LAC becomes comparable to the hyperfine coupling strength. As an application of the nuclear spin manipulation, we show nuclear magnetic resonance (NMR) and coherent manipulation of the nuclear spin via Rabi oscillations (Fig. 5 d,e). For the measurement of a nuclear magnetic resonance spectrum, we apply a radio-frequency (RF) π-pulse of variable frequency. We observe a resonance dip at 450(1) kHz, corresponding to the Larmor precession frequency $\nu_{13C} = \gamma_{13C} B = 447\ 5$ kHz of a bare $^{13}$C nucleus. Here, $\gamma_{13C} = 10.705$ kHz/mT is the gyromagnetic ratio of the $^{13}$C nuclear spin and $B = 41.8(5)$ mT is the applied field (corresponding to the LAC). We can thereby identify the coupled nuclear spin as a $^{13}$C lattice spin. Note the absence of a chemical shift which indicates that indeed the electron spin is absent. For the measurement of nuclear Rabi oscillations, we apply a resonant RF pulse of variable length between the initialization and readout laser pulses. Several Rabi oscillations are observed that decay on a time scale of about 1 ms. The decay is likely to originate from dipole-dipole coupling of the nuclear spin to surrounding $^{13}$C lattice spins. Note that the dominant decay mechanism is spin diffusion. By contrast to ensemble magnetization (that decays with $T_1$), here the relevant time scale is the ensemble $T_2^*$ time that is about 10 ms in diamond with natural $^{13}$C abundance[22] and agrees reasonably with the observed decoherence time. For future increase of the coherence time, such dipole interactions between nuclei of the same species can be efficiently removed by homonuclear decoupling schemes[9,23–25]. Other sources of decoherence could be paramagnetic or nuclear spin impurities at the surface whose decoherence effects could be avoided by isolating ST1 defects in the bulk crystal.

The defect spin system presented in this work holds promise for room temperature solid state nuclear spin memories and quantum registers with ultra-long spin coherence times. Such a device could be used as a memory in integrated hybrid quantum systems[13,14] or as a solid state quantum token[9]. For such applications, coherence times on the order of seconds or hours are desirable. Owing to its spin-free ground state, the system presented here is a promising candidate for achieving such long coherence times by homonuclear dynamical decoupling, such as BLEW[23], FSLG[24], MREV[25]. An increase of the nuclear spin coherence time by several orders of magnitude is expected[9]. In diamond, the coherence

time of $^{13}$C nuclear spins (at room temperature) is ultimately limited to time scales on the order of one day[22].

**Methods**

Polarization and readout of the triplet spin state

A long intense laser pulse is applied to prepare the system in a spin-polarized state of the triplet level. The intensity and length of the pulse are chosen such that the optical transition is saturated and steady state fluorescence is reached during the pulse. Thereby the majority of the population is transferred to the slowly decaying triplet sublevel, while the two other triplet levels and the ground and excited states are nearly empty. The laser pulse is followed by a short idle time (~50 ns) to allow any residual population of the excited state to decay.

To readout the triplet spin state, the fluorescence is recorded during a laser pulse that is identical to the preparation pulse. Decaying fluorescence response curves are observed that encode the triplet populations. In particular, the fluorescence counts summed over a window at the beginning of the response profiles are proportional to the three triplet spin populations. This fluorescence is plotted in Fig. 3 and 4. In all data shown, a 200 ns integration window was used.

Polarization and readout of the nuclear spin

The nuclear spin is initialized by applying a long (10 µs) laser pulse (duration longer than several triplet lifetimes) followed by an idle time (20 µs) that ensures that the triplet population is fully decayed to the electronic ground state prior to any nuclear spin manipulation. The readout consists of a second laser pulse with simultaneous detection of the fluorescence, as in the case of the electron spin manipulation, except that now a longer (8 µs) integration window is used to accumulate fluorescence contrast. As before, the integrated fluorescence encodes the spin state.

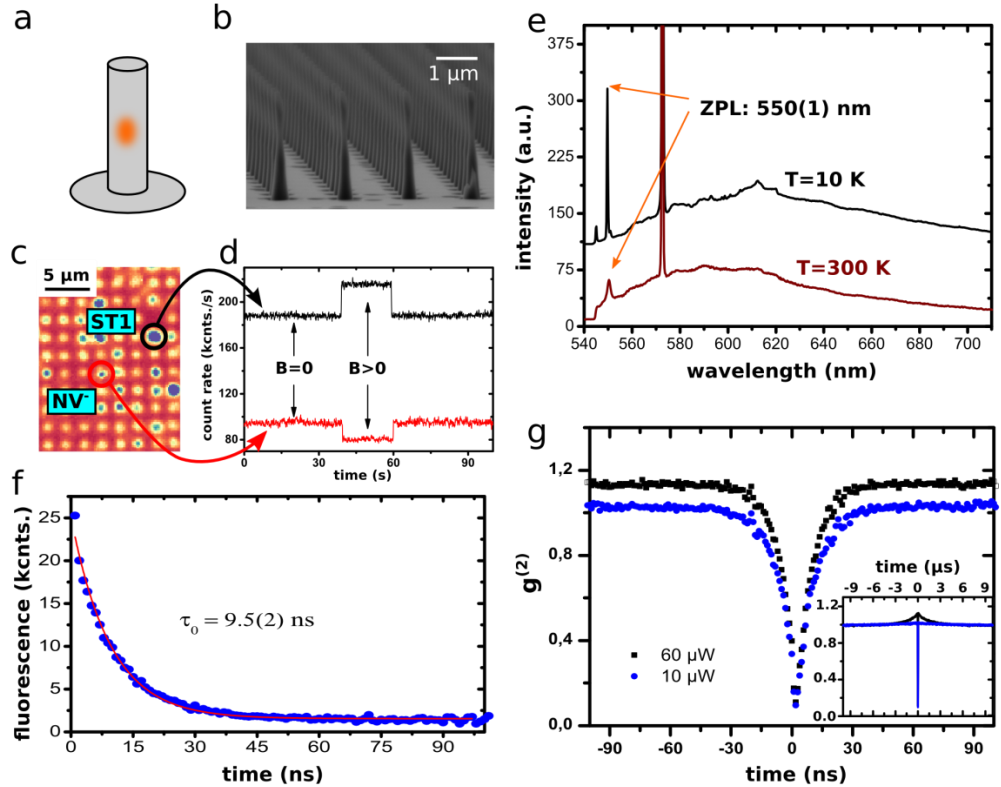

**Fig. 1| Optical properties of the ST1 defect. a,** Illustration of a diamond nano wire containing a single fluorescent point defect. **b,** SEM image of the [111] HPHT diamond sample with fabricated nano wires (diameter ~200 nm, length ~2 μm). **c,** confocal fluorescence microscope image of the sample. The marked wires contain single NV- and ST1 defects. **d,** fluorescence time trace recorded during application and removal of a non-aligned magnetic field. The fluorescence is quenched and enhanced for NV-, and ST1 defects, respectively. **e,** fluorescence emission spectrum of a single ST1 defect at room temperature and liquid helium temperature. A zero phonon line at 550(1) nm and a broad emission band in the red to NIR range is observed, **f,** fluorescence decay showing an excited state life time of 9.5 ns, **g,** second order photon correlation at increasing excitation power. The dip at delay t=0 reaches well below 0.5, proving the single center nature of the defect. At high excitation power pronounced bunching shoulders are observed, that indicate the presence of a long lived shelving state.

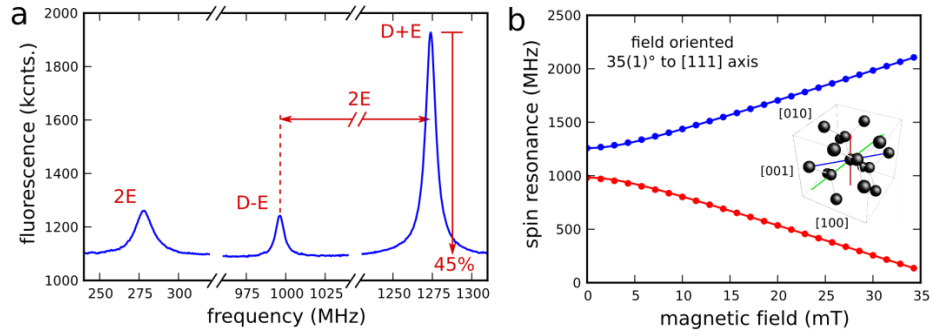

**Fig. 2| Optically detected electron spin resonance. a,** room temperature ESR spectrum at zero magnetic field. Resonant transitions are indicated by their zero-field splitting parameters D=1135(1) MHz and E=139(1) MHz. We observe an exceptionally high optical spin signal contrast of the D+E spin projection of about 45 %, **b,** Zeeman curve with an aligned axial magnetic field oriented along the defect main symmetry axis (35(1)° w.r.t. the [111] crystal axis); dots: experiment, lines: theoretical fit obtained from the triplet spin Hamiltonian $H = D \left[S_z - S(S+1)/3\right] + E \left[S_x - S_y\right] + g S_z B_z$, where $S_\alpha$ are the usual spin operators for total Spin $S = 1$, $g$ is the $g$-factor and $B_z$ is the axial field (see SI). The inset shows a schematic illustration of the crystallographic orientation of the defect. Green, blue and red arrows denote the [111], [110] (defect's main axis) and [001] ($C_{2v}$ symmetry axis), respectively.

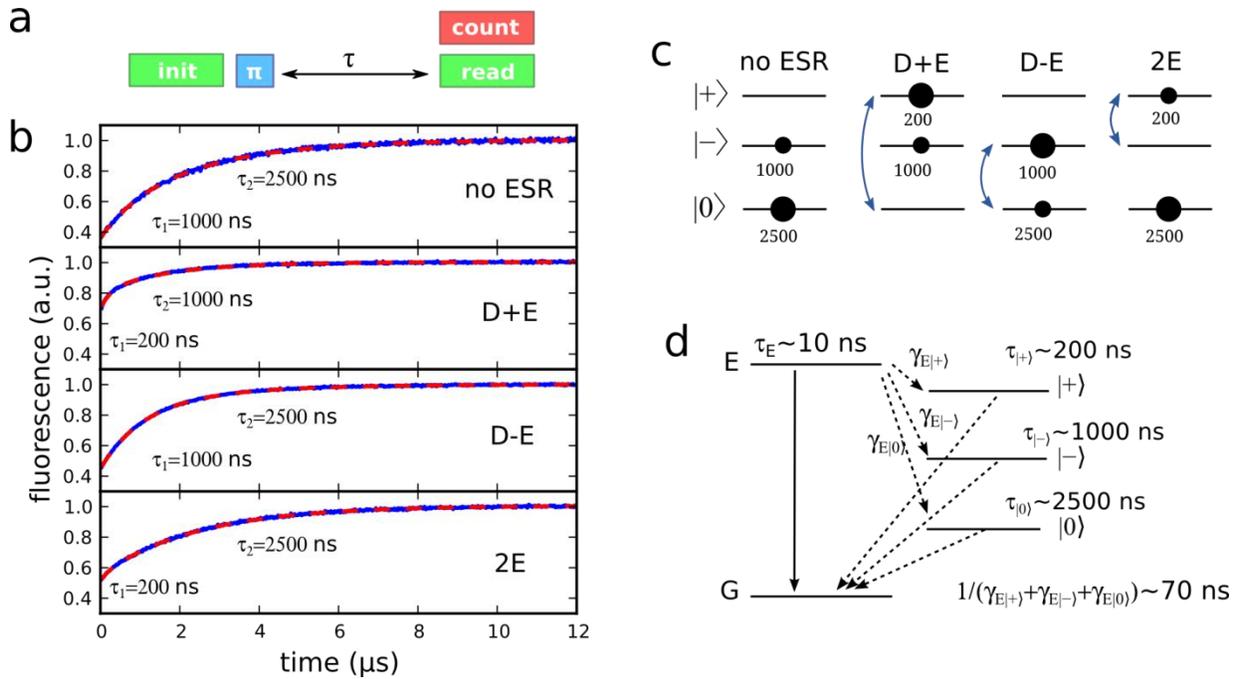

**Fig. 3 | Kinetics of the shelving state and energy level scheme. a,** optical and microwave pulse sequence for the measurement of the shelving state lifetimes. After an optical pump pulse, a resonant microwave π-pulse is applied to exchange the populations of selected spin sublevels and thereby control the dominant decay constants, **b,** Decay curves of the shelving state population described by double exponential functions $y = c - a_1 \exp(-t/\tau_1) - a_2 \exp(-t/\tau_2)$. Blue: experimental data, red dashes: double exponential fits (see SI). For each panel, the microwave π-pulse is resonant with the denoted ESR transition, **c,** Assignment of the decay constants to the triplet energy levels. The initial populations (no ESR), are proportional to the decay constants. The small population of the short lived state is omitted. The observed decay constants imply the assignment of the energy levels denoted on the left, **d,** Summary of the triplet kinetics and electronic structure consisting of singlet ground- and excited states and a meta-stable triplet state. Transition rates between two states $i$ and $j$ and the lifetime of a state $i$

are denoted $\gamma_{ij}$ and $\tau_i$, respectively.

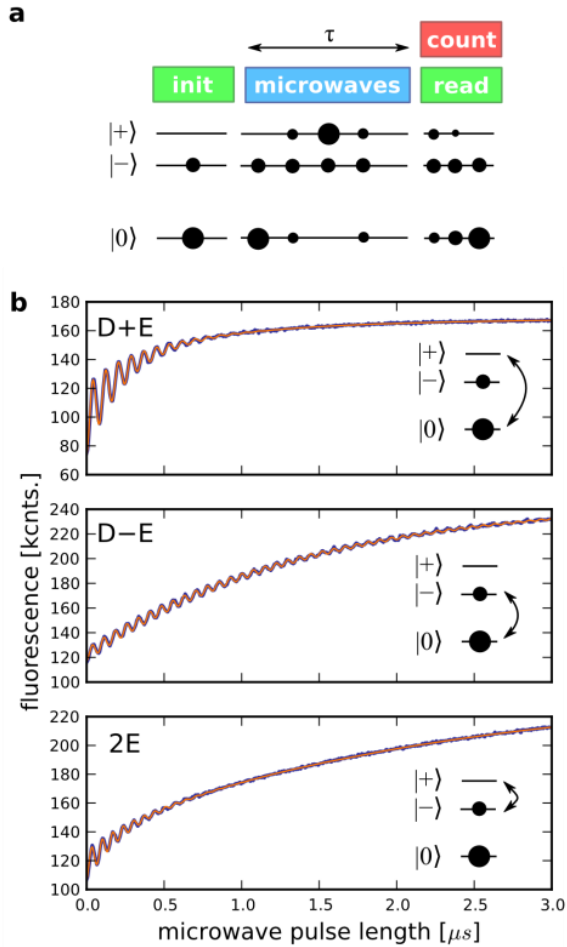

**Fig. 4| Coherent triplet spin manipulation. a,** Optical and microwave pulse sequence and evolution of the triplet populations (here illustrated for microwaves resonant with the D+E transition), **b,** Rabi oscillations among the triplet spin sublevels. Thick blue lines show experimental data, thin red lines show theoretical fits with the model $I = I_0 - \left(a_1 + a_2 \cos \Omega t\right) \exp{-t/\tau} - a_3 \exp(t/\tau)$, where $\tau = 2(\tau_\alpha^{-1} + \tau_\beta^{-1})^{-1}$ is a combined decay constant of the levels $\alpha$ and $\beta$ involved in the Rabi oscillations, $\tau$ is the decay constant of the remaining level, $a_i$ are fluorescence amplitudes and $\Omega$ is the Rabi frequency. Note that for the upper and lower panel the combined decay constant is dominated by the fast decay of the $|+\rangle$ level. For the fit shown in the middle panel the last term is omitted, since the unaffected level is the $|+\rangle$ level, which (nearly) unpopulated.

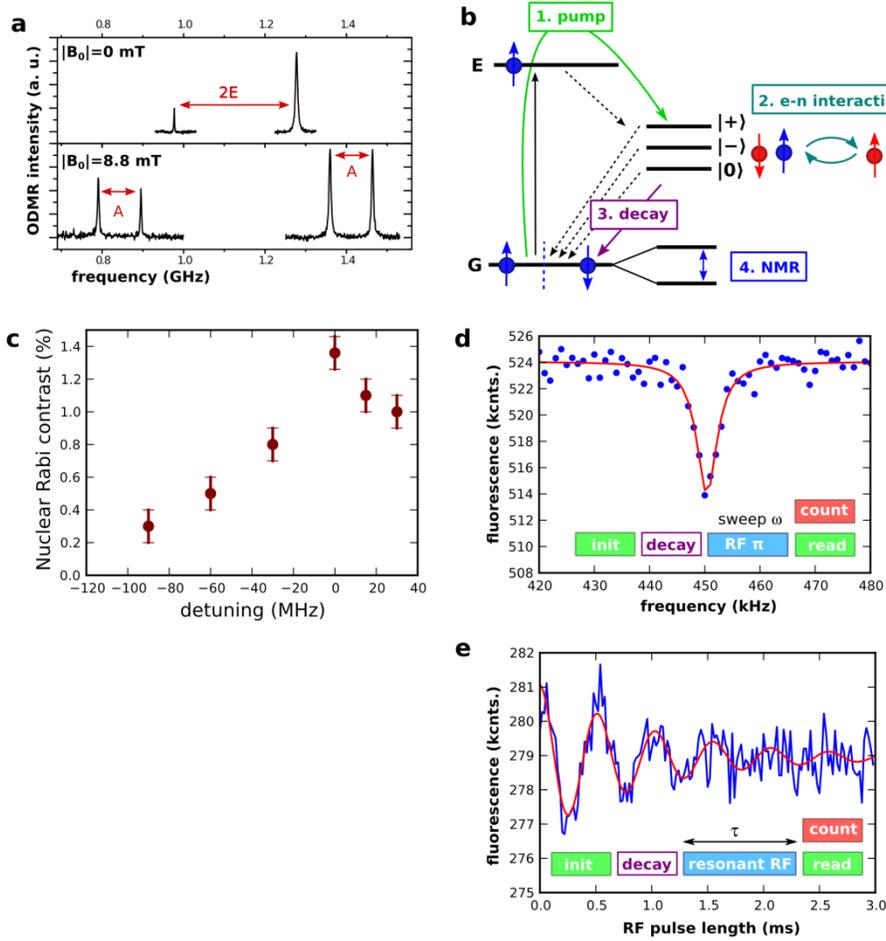

Fig. 5 | **Readout and control of a single nuclear spin in the spin-free electronic ground state. a,** ODMR spectra of an ST1 defect with a nearby $^{13}$C nuclear spin, under zero (upper panel) and moderate (lower panel) magnetic field applied along the defect's axis. E and A denote the zero-field- and hyperfine-splitting parameters, respectively. The hyperfine splitting is suppressed at small fields and becomes visible when the Zeeman energy is larger than the E parameter. **b,** Readout and control of the nuclear spin with the electronic triplet spin ancilla. The nuclear spin (blue arrows) is interrogated by optically pumping the electron spin (red arrows) into the triplet state and allowing the electron and nuclear spin to interact. The subsequent decay of the triplet state destroys the electron spin and leaves the nuclear spin unchanged. The bare nuclear spin is then coherently manipulated by NMR. Efficient transfer of spin polarization between the electron and nuclear spin in the triplet state is provided by exploiting an ONP as a consequence of LAC between the $|-1,+1/2\rangle$ and $|0,-1/2\rangle$ hyperfine levels (see text), **c,** Signal contrast of nuclear Rabi oscillations, corresponding to the product of the nuclear spin polarization and readout contrast, as a function of detuning from the LAC (zero detuning corresponds to the LAC). **d,e,** NMR spectrum and Rabi oscillations of the nuclear spin in the electronic ground state. For the NMR spectrum (d), an RF π-pulse of variable frequency is inserted between the preparation and readout laser pulses. The resonance dip corresponds to the Larmor precession of a bare $^{13}$C nuclear spin. Blue: experimental data, Red: Lorentzian fit. For the Rabi oscillations (e) a resonant RF pulse of variable length

is inserted between the initialization and readout laser pulses. Blue: experimental data, red: exponentially decaying cosine fit; decay constant 1.0(1) ms.


\title{Readout and control of a single nuclear spin with a meta-stable electron spin ancilla\\
--\\
Supplementary information}
\author{S.--Y. Lee, M. Widmann, T. Rendler, M. Doherty, T. Babinec, S. Yang,\\ M. Eyer, P. Siyushev, B. Haussmann, M.
Loncar, Z. Bodrog,\\ A. Gali, N. Manson, H. Fedder, J. Wrachtrup}
\maketitle

\section{Introduction}

Here we present supporting material to the key results and arguments found in the main text. In particular, we analyze
the properties of the ST1 defect that we observed through our employment of various optical spectroscopy and magnetic
resonance techniques. From our analysis, we draw key conclusions about the structure of the ST1 defect and its
electronic levels. These conclusions will provide the foundations for future studies of the formation of the ST1 defect
and its properties. Our methodology draws many parallels with the vital early investigations of the NV center
\cite{Davies1974,Davies1976,Loubser1978,Collins1983,He1993,Gruber1997,Drabenstedt1999}. Further, we consider the
mechanisms underlying optical nuclear spin-polarization, -readout and -decoherence in greater detail. Here we draw on
early works on optical nuclear polarization \cite{Stehlik1975,Colpa1977} and $^{13}$C relaxation in diamond
\cite{Reynhardt1997,Reynhardt2001}.

\section{Sample fabrication and the occurrence of ST1 defects}

Arrays of diamond nano wires (length 2~$µ$m, diameter 200~nm, pitch 2~$µ$m) were fabricated into the cleaved [111]
surface of an HPHT diamond (element six Ltd.) by reactive ion etching as described in detail in \cite{Hausmann2010}. In
brief: first, an HSQ mask is defined with e-beam lithography and a lift-off
process. Thereafter, anisotropic etching is performed with an inductively coupled oxygen plasma. After the nano wire
fabrication, the sample is annealed in vacuum at 850~$^\circ$C for several hours and subsequently boiled in a mixture of
equal amounts of perchloric-, sulfuric-, and nitric acid at 100~$^\circ$C. 32 arrays, each containing 80$×$80 nano
wires were fabricated and the ST1 defects were found in all patches that were investigated. A randomly chosen area
(100$µ$m$×$100$µ$m) on one of the patches was intensively investigated in a home built confocal fluorescence
microscope (Nikon LU Plan Fluor Epi 100x, 0.9 N.A. air objective, 532 nm laser excitation) and 19 single ST1 defects
were found within this area. This number is equivalent to approximately one ST1 defect per 800 nano-wires. Note that the
number of NV defects in this area is considerably smaller than the number of ST1 defects (only about 5) and
agrees roughly with the amount expected from the bulk NV- concentration. All ST1 defect fluorescence was observed on
nano-wires. The bulk diamond underneath and the surface next to the wires was carefully investigated by fluorescence
microscopy and no ST1 defects were observed. Note that this fact might be biased by the positive role of the nano-wires
on the optical collection efficiency \cite{Babinec2010}. Due to the high collection efficiency of the nano wires, ST1
defects in the bulk would appear a factor of about 5 times dimmer than the defects on the nano-wires. This might render
them hardly visible against the background fluorescence, which is about 10 kcnts./s on the plane sample surface and
about 100 kcnts./s on an empty wire the present case. Since the nano-wires have a tendency to attract
fluorescent material from the environment whose emission then couples efficiently into the wires and produces background
fluorescence, during the past, the sample surface had been cleaned several times by slowly scanning the laser over the
surface (incident power $\sim$10 mW). This procedure bleaches most of the adsorbed molecules and removes most of the
background fluorescence. At present we cannot exclude that this strong laser illumination in combination with ambient
adsorbates plays a role for the occurrence of the ST1 defects. Thus, we can identify
three aspects that might contribute to the occurence of ST1 defects.
\begin{itemize}
\item The [111] crystal surface
\item The fabrication process, in particular the plasma etching using oxygen gas (trace amounts of
contaminats from the chamber and electrodes may also play a role) and subsequent annealing
\item Adsorbate molecules in combination with strong 532 nm laser illumination
\end{itemize}
As we will discuss below, oxygen is a possible defect constituent since it does not have a nuclear spin and is capable
of $sp3$ bonding. Future oxygen implantation and reactive ion etching studies can be used to probe this hypothesis.
Recently a defect with a ZPL at 547.2~nm has been reported in HPHT diamonds grown with a
Ni-containing catalyst at elevated temperatures (>1400~$^\circ C$); PL spectra measured under 337.1~nm excitation
\cite{Iakoubovskii2002}. We think that the difference of the observed ZPL compared to the ZPL observed in the present
work at 550(1)~nm is too large to identify both defects as the same species. Nevertheless, a shift of a ZPL e.g. due to
lattice deformations or co-presence of another impurity species might still be possible. More importantly, for the
547.2~nm defect, an excited state lifetime of 150~$µ$s was reported which differs substantially from the
excited state lifetime of the ST1 defect of about $10$~ns. Our current working hypothesis is that the defect observed in
the present work is a different one.

\section{Properties of the optical fluorescence}

\begin{figure}
\centering\includegraphics[width=0.75\columnwidth]{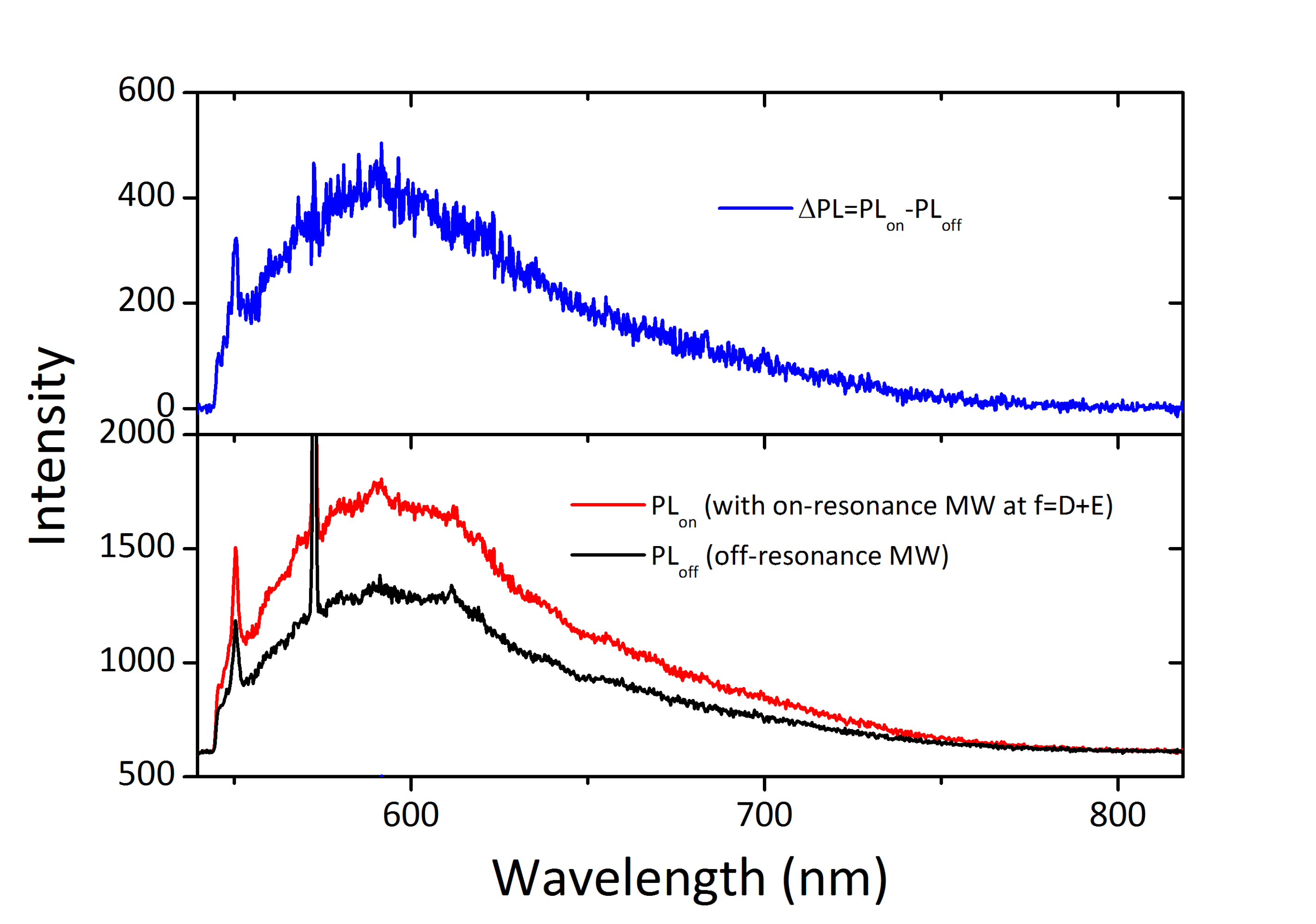}
\caption{\label{fig:PL}\textbf{Enhancement of emission induced by ESR}. Low panel: Two PL spectra measured at room
temperature with on- and off-resonance continuous ESR excitation. Upper panel: Difference between two spectra.
Enhancement of both ZPL and phonon side band can be seen. The sharp peak at 573 nm is the Raman peak.}
\end{figure}

Under 532 nm excitation, the fluorescence of the ST1 defect exhibits a zero phonon line (ZPL) at 550 nm (2.25 eV) and a
broad vibronic band that extends to lower energy $\sim750$ nm (refer to figure \ref{fig:PL}). The association of this
fluorescence with the ST1 defect (and not with some other unidentified emission states in the nano-wires) was
established by two different methods. Firstly, in order to verify that this emission is responsible for the observed
single photon emission, the second order correlation measurement was repeated in the presence of a band pass
filter centered at 550 nm (band width 25 nm) and identical anti-bunching behavior was observed. In order
to confirm that ESR gives rise to the enhancement of this emission, the PL spectra were measured while applying
on- and off-resonant microwave excitation. Figure~\ref{fig:PL} shows that both the ZPL and its phonon side band were
enhanced when the on-resonance microwave was applied. Note that the PL without MW was identical to PL with
off-resonance MW (not shown).

\begin{figure}
\centering\includegraphics[width=0.375\columnwidth]{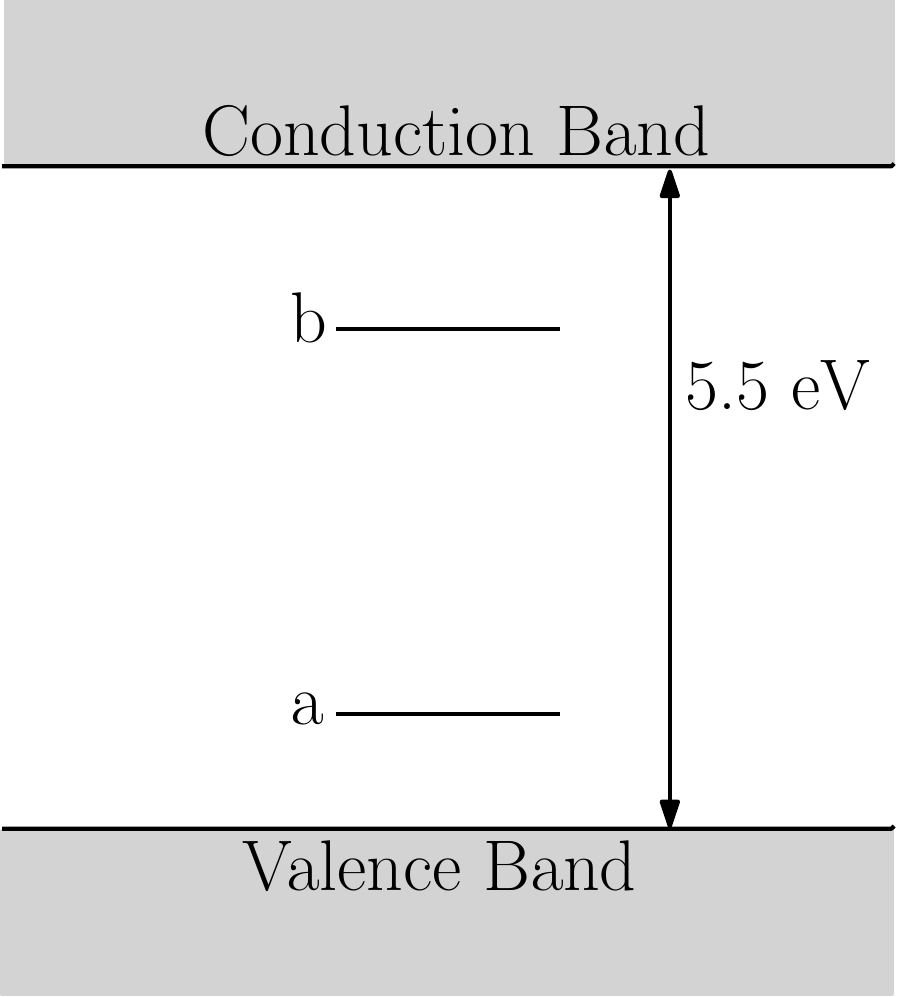}
\caption{\label{fig:bandstructure}\textbf{Electron orbitals of the ST1 defect}. The shaded regions represent the diamond
valence and conduction bands separated by the indirect bandgap of $\sim5.5$ eV. The solid lines within the band gap
represent the energies of the orbitals of the ST1 defect that are involved in the optical transition. The $a$ orbital is
the highest occupied orbital in the electronic ground state. Optical excitation results in the promotion of an electron
from the $a$ orbital to the lowest unoccupied orbital $b$. The symmetries of the orbitals are unknown. There may exist
other orbitals of the ST1 defect in the bandgap.}
\end{figure}

The energy of the fluorescence and the existence of the sharp ZPL implies that the electronic transition occurs between
energetically discrete electron orbitals within the diamond bandgap (refer to figure \ref{fig:bandstructure})
\cite{stoneham2001theory}. As the shape of the vibronic band in the PL spectra of figure \ref{fig:PL} is distorted by
the first
and second order Raman resonances of the 532 nm excitation, it is not possible to directly establish the Huang-Rhys
factor of the fluorescence. The Huang-Rhys factor describes the relative intensity of the ZPL and the vibronic band and
is a measure of the lattice relaxation associated with the electronic transition \cite{stoneham2001theory}. Since the
Raman
resonances should not change in the presence of microwaves, the difference in the PL spectra while applying on- and
off-resonance microwaves provides a reasonable picture of the ZPL and vibronic band that is free of distortions from the
Raman resonances. It is apparent that like many other color centers in diamond, the
Huang-Rhys factor at room temperature is appreciable (estimate at least greater than one) and that there is an
appreciable lattice relaxation associated with the electronic transition \cite{Davies1974,Davies1981}. The difference
between the first moment of the vibronic band and the ZPL energy is $\sim 0.15$ eV at room temperature, which provides
an estimate of the Anti-Stokes shift and lattice relaxation energy associated with the transition
\cite{stoneham2001theory}.
Further studies of the vibronic band will provide additional information on the electron-phonon interactions of the ST1
defect \cite{stevenson1966phonons}.

\section{Electronic structure and optical cycle}
\label{sec:electronic_structure}

It was discussed in the main text that the key experimental observations that suggest the existence of
a metastable triplet state are the {\em increase of the fluorescence} under application of a
strong misaligned magnetic field or resonant microwave field, and {\em pronounced photon
bunching} under saturated optical excitation (refer to Fig.~1, main text).
Such behaviour is well known for fluorescent molecules with
singlet ground and excited states and a long lived excited triplet state that
features different lifetimes of the individual triplet spin sublevels~\cite{Chiha1978}.
The triplet state plays the role of a shelving state. When the population is cycled between the
ground and excited state by optical pumping, part of the population gets
trapped in the triplet state. This Inter-System-Crossing (ISC) leads to an overall reduced fluorescence since the
ISC transitions are typically non-radiative and the lifetime of the triplet
state is long (ISC transitions are only weakly allowed by spin-orbit or
spin-spin interaction). The long lifetime leads to characteristic 'bunching
shoulders' in the second order photon correlation. When either the population- or the decay rates of the triplet
sublevels are different, the sublevels are not equally populated, i.e., the spin
becomes polarized. Spin polarization is typically built up during several
optical pump cycles. When the decay rates of the triplet sublevels are
different, a mixing of the triplet populations (either by a resonant microwave
field or by a strong misaligned magnetic field), results in a faster or slower
deshelving of the triplet state. This leads to an increase or decrease of the emission
intensity.

In the following, we analyze the optical cycle and provide conclusions about the defect electronic
structure. We first present a 5--level rate equation model to describe the optical cycle. Thereafer we
experimentally determine the transition rates. Finally we draw conclusions for the electronic structure and
symmetry of the electronic states.

\subsection{Rate equation model}

\begin{figure}
\centering\includegraphics{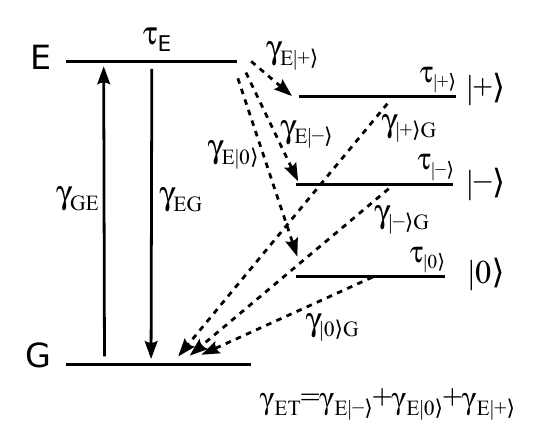}
\caption{\label{fig:rate_model}\textbf{Rate equation model.} The ground and excited states are denoted $G$ and $E$. The
three triplet spin levels are denoted $|0\rangle$, $|-\rangle$ and $|+\rangle$. Rates and lifetimes are denoted by
greek letters $\gamma$ and $\tau$, respectively.}
\end{figure}

The rate equation model is illustated schematically in figure~\ref{fig:rate_model}. It consists of five levels: the
ground and excited states $G$ and $E$ and the three triplet spin levels
$|0\rangle$, $|-\rangle$ and $|+\rangle$. Rates between levels $µ$ and $\nu$ are
denoted by $\gamma_{µ\nu}$ and the lifetime of a levels $µ$ is denoted by
$\tau_{µ}$. $\gamma_{GE}$ is the optical pump rate, $\gamma_{EG}$ is the spontaneous emission rate,
$\gamma_{E\nu}$, $\nu=|0\rangle,|-\rangle,|+\rangle$ are the triplet population rates and $\gamma_{µG}$,
$µ=|0\rangle,|-\rangle,|+\rangle$ are the triplet decay rates.
The rates are related to the lifetimes $\tau_µ=1/\sum_{µ}\gamma_{µ\nu}$.
The time evolution of the populations $n=(n_{G},n_{E},n_{|+\rangle},n_{|-\rangle},n_{|0\rangle})$ is given by the
rate equations
\begin{equation}\label{eq:rate_eq}
\begin{aligned}
\dot{n}_{G}=&-\gamma_{GE}n_G+\gamma_{EG}n_E+\gamma_{|+\rangle G}n_{|+\rangle}+\gamma_{|-\rangle
G}n_{|-\rangle}+\gamma_{|0\rangle G}n_{|0\rangle}\\
\dot{n}_E=&+\gamma_{GE}n_G-\gamma_{EG}n_E-(\gamma_{E|+\rangle}+\gamma_{E|-\rangle}+\gamma_{E|0\rangle})n_E\\
\dot{n}_{|+\rangle}=&+\gamma_{E|+\rangle}n_E - \gamma_{|+\rangle G}n_{|+\rangle}\\
\dot{n}_{|-\rangle}=&+\gamma_{E|-\rangle}n_E - \gamma_{|-\rangle G}n_{|-\rangle}\\
\dot{n}_{|0\rangle}=&+\gamma_{E|0\rangle}n_E - \gamma_{|0\rangle G}n_{|0\rangle}\\
\end{aligned}
\end{equation}
Using the conservation of the total population
\begin{equation}
 1=n_G+n_E+n_{|+\rangle}+n_{|-\rangle}+n_{|0\rangle},
\end{equation}
one euqation and one variable in Eq.(\ref{eq:rate_eq}) can be eliminated. The result is a regular first order
linear differential equation with constant coefficient matrix $A$ and right hand side vector $b$.
\begin{equation}
  \dot{n}=A n + b,
\end{equation}
which has the solution
\begin{equation}
 n(t)=e^{At}n_0+A^{-1}e^{At}b - A^{-1}b,
\end{equation}
where $n_0$ are the initial populations. Using such a model, all photophysical properties, such as the saturation
behavior, anti-bunching, fluorescence response to an optical pump pulse, etc., can be easily computed numerically by
diagonalizing the coefficient matrix $A$ or in some cases analytically. In the following
sections, we will experimentally determine the decay rates $\gamma_{|+\rangle G}, \gamma_{|-\rangle G},
\gamma_{|0\rangle G}$ and the total population rate
$\gamma_{ET}=\gamma_{E|+\rangle}+\gamma_{E|-\rangle}+\gamma_{E|0\rangle}$ of the triplet level, where we
assume that the population rates of the triplet state are faster than the decay rates and are equal for all spin
sublevels, and the triplet decay rates are distinct. In this case, the populations of the triplet levels under
saturated optical excitation are proportional to their lifetimes.

\subsection{Triplet lifetimes}
\begin{figure}
\centering\includegraphics{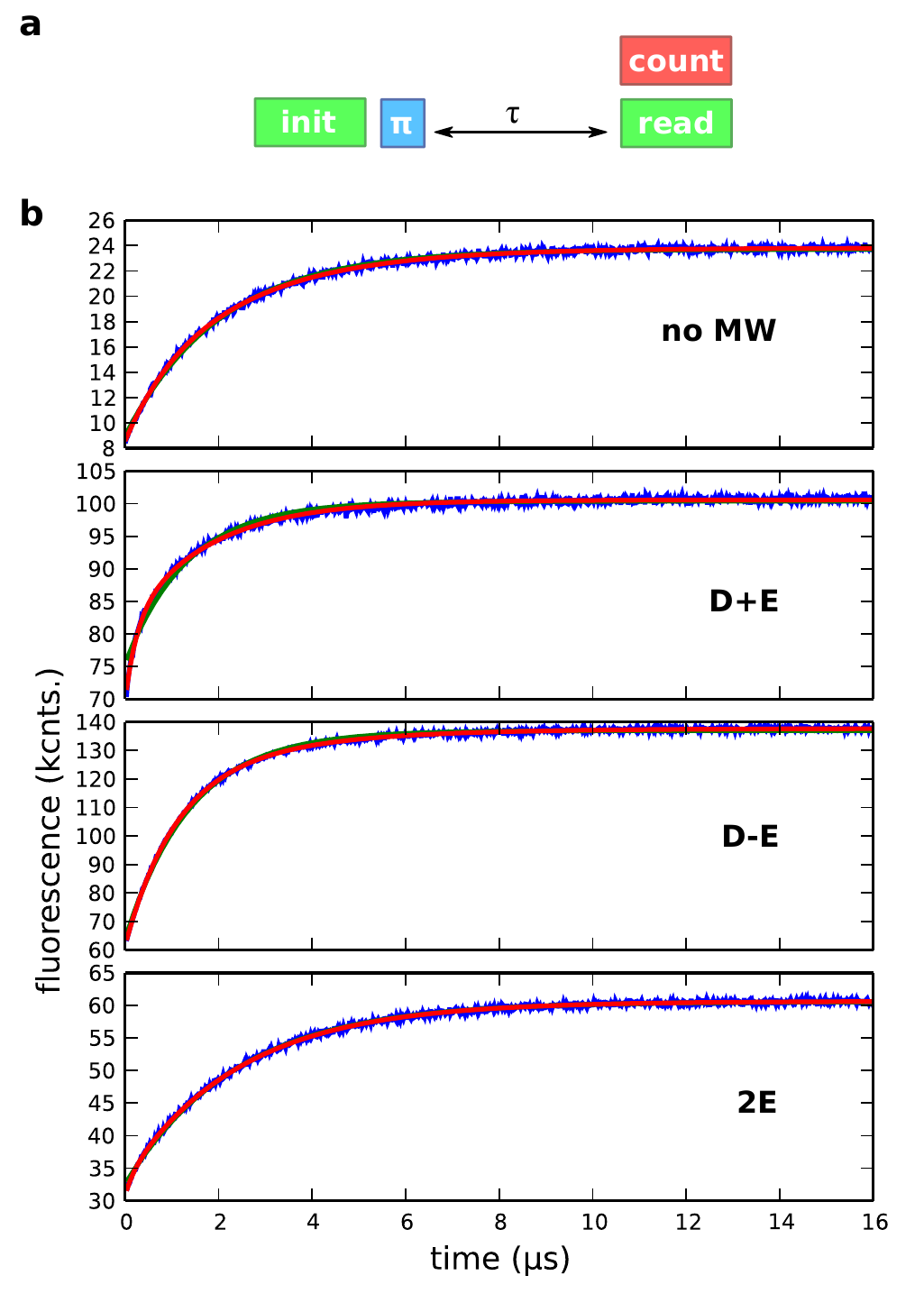}
\caption{\label{fig:triplet_lifetimes}\textbf{Experimental determination of the Triplet lifetimes.} \textbf{a,}
Pulse sequence. \textbf{b,} Measured decay curves for defect C (see text),
without microwave pulse and with a $\pi$--pulse tuned in resonance with the
three ESR transitions. Blue dots: experimental data, solid green lines: single
exponential fits, solid red lines: double exponential fits.}
\end{figure}
As discussed in the main text, in order to determine the triplet decay rates, the recovery of the ground state
population was observed by a pulsed optical experiment (refer to Fig.3, main text and
figure~\ref{fig:triplet_lifetimes}). First a long laser pulse is applied (duration longer than the triplet lifetimes,
power high enough to saturate the optical transition). At the end of the
pulse the system is in the steady state under saturated optical illumination, i.e., the triplet levels are maximally
populated. The relative populations of the sublevels are
proportional to their lifetimes. Next, the triplet levels are
allowed to decay for a variable amount of time in the range few ns to several $µ$s. Finally, the population of
the ground state is measured by applying a second laser pulse that is identical to the first pulse and
simultaneously recording the fluorescence. The integrated fluorescence is proportional to the ground state population
before the pulse. Most data was accumulated until the standard deviation of the measured ground state populations (given
by photon shot noise) was below 1\%. Exponentially increasing fluorescence curves were obtained.
The time constants are given by the lifetimes of the triplet sublevels. The data was fit with single-, double- and
triple-exponential models and the goodness-of-fit $Q=\Gamma(0.5\nu,0.5\chi^2_0)$ was evaluated for
each fit~\cite{press1992numerical}. Here $\Gamma$ is the complementary incomplete gamma function,
$\nu$ is the degrees of freedom and $\chi^2_0$ is the $\chi^2$ evaluated at the minimum. Single exponential fits
typically result in goodness of fit values $Q<10^{-10}$ and can be discarded. Most Double-
and triple-exponential fits result in goodness-of-fit values in the range $1<Q<10^{-3}$ and are both
acceptable. The
triple-exponential models typically contain one component whose amplitude is much smaller than the two other components
and therefore this component can be safely neglected. This reflects the fact that acceptable goodness-of-fit
values are readily obtained with double exponential fits. We conclude that double exponential fits are faithful
models for all data.
To assign the lifetimes to the spin sublevels, we perform the measurement with an additional microwave $\pi$-pulse
inserted after the optical pump pulse. The microwave power is high enough, such that the duration of the
microwave pulse is short compaired to the lifetimes. The microwave frequency is tuned in resonance with each of the
triplet spin transitions. The microwave $\pi$-pulse swaps the population of the triplet sublevels and thereby
exchanges the amplitudes of the corresponding exponential decay curves. Table~\ref{tab:triplet_lifetimes} summarizes
the measured decay constants and amplitudes and specifies the goodness-of-fit values for three different ST1 defects.
Importantly, there are only minor deviations among the different defects. Generally three distinct time
constants are observed, of about $200$~ns, $1000$~ns and $2500$~ns. The result is summarized in the last row. The
result is used to assign the time constants to the spin levels as shown in the illustration below the table. To obtain
an overall average for the decay constants, we evaluate the mean and standard deviation over all equivalent time
constants, yielding $\tau_{|0\rangle}=2500(204)$ ns, $\tau_{|-\rangle}=1072(350)$ ns,
$\tau_{|+\rangle}=209(44)$ ns.
\begin{table}
\begin{center}
 \begin{tabular}{c|rrrr}
 \toprule
  ST1 & no MW & D+E & D-E & 2E\\
  \hline
   &930(37) & 161(5) & 775(39) & 158(19) \\
   &2288(21) & 1029(13) & 2358(32) & 2448(13) \\
  A&11054(700) & 37492(695) & 8212(515) & 1249(20) \\
   &41878(720) & 44867(672) & 22750(532) & 11988(40) \\
   &4.7e-2 & 3.3e-2 & 4.2e-2 & 6.0e-2 \\
  \hline
   &761(52) & 267(7) & 865(9) & 239(32) \\
   &2329(35) & 1641(24) & 2908(74) & 2390(9) \\
  B&3630(316) & 23878(340) & 60634(694) & 2068(136) \\
   &13419(323) & 20388(336) & 19072(712) & 32700(93) \\
   &4e-5 & 2e-32 & 7e-5 & 2.6e-3 \\
  \hline
   &936(74) & 247(9) & 1055(15) & 181(20) \\
   &2473(56) & 1757(17) & 2812(99) & 2447(9) \\
  C&3838(493) & 11730(216) & 56834(1256) & 2444(147) \\
   &11396(499) & 19118(194) & 19265(1272) & 27342(69) \\
   &2.3e-3 & 1.5e-13 & 1.3e-3 & 2.2e-2 \\
  \hline
          & 1000 & 200 & 1000 & 200 \\
  summary & 2500 & 1000 & 2500 & 2500 \\
  \bottomrule
 \end{tabular}
 \rule{8mm}{0mm}\includegraphics{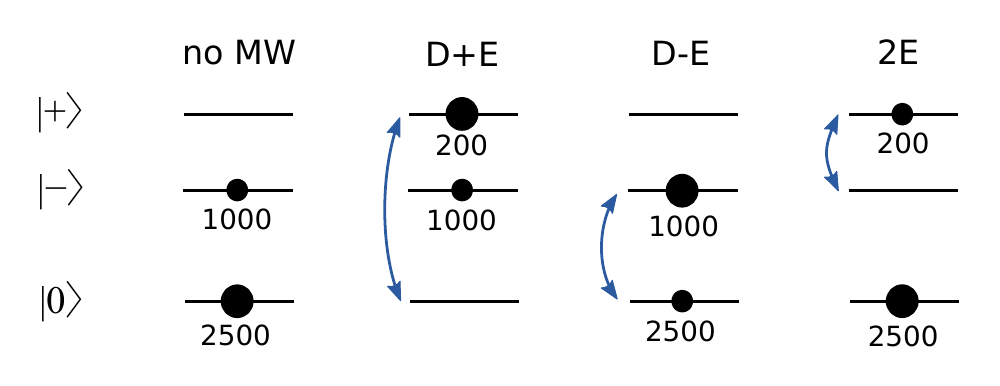}
\end{center}
\caption{\label{tab:triplet_lifetimes}
\textbf{Measured triplet lifetimes of three ST1 defects A,B,C.} The data in each cell shows (from top to bottom) the two
time constants [ns], the two amplitudes [counts] and the goodness-of-fit of a double exponential model. The
last row summarizes the observed time constants [ns]. The mean and standard deviation over equivalent time
constants yields $\tau_{|0\rangle}=2500(204)$ ns, $\tau_{|-\rangle}=1072(350)$ ns, $\tau_{|+\rangle}=209(44)$ ns}.
\end{table}

\subsection{Triplet population rates}

\begin{figure}
\centering\includegraphics[width=1\columnwidth]{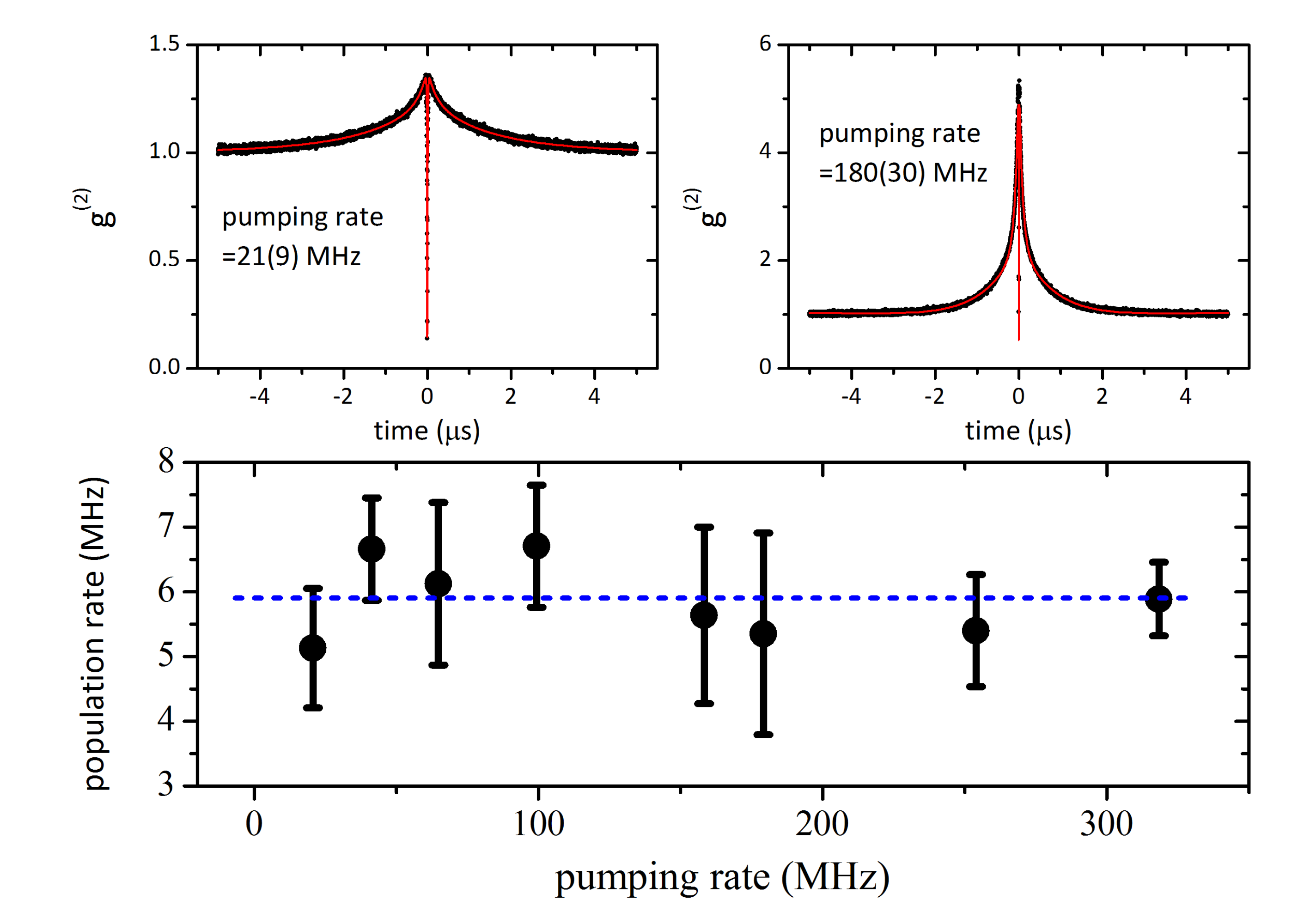}
\caption{\label{fig:pop_rate}\textbf{Determination of the population Rates.} Two upper panel: measured $g^{(2)}$
functions at low and high optical excitation powers. Red lines are the fitted curves. Instead of the excitation powers,
the extracted pumping rates are shown. Lower panel: The obtained population rates as a function of the pumping rate.
Only a common value 5.9(6) MHz was obtained because the three population rates could not be distinguished. Blue dotted
line indicates this value which is almost constant.}
\end{figure}
In the following, we determine the triplet population rates by two different methods.
One approach exploits the second order correlations, $g^{(2)}$, of the photon
emission at different excitation powers. The numerical solutions is fit based on the five level rate model
described in equation~(\ref{eq:rate_eq}) yielding the population rate parameters. The measured
$g^{(2)}$ functions and fitting results are shown in
figure~\ref{fig:pop_rate}. Two representative $g^{(2)}$ functions measured at low and high excitations are shown. Both
the anti-bunching part and bunching shoulder can be well identified. Since the excited singlet state life time and the
triplet state life times are already known, the measured $g^{(2)}$ functions can be fitted with four free
parameters, namely the pump rate and the three population rates. Three population rates, however, could not be
distinguished within the error range, instead the fit was simplified to two free parameters by considering the
total rate $\gamma_{ET}=(56(7)\,\textrm{ns})^{-1}$, where we
assume $\gamma_{E|+\rangle}=\gamma_{E|-\rangle}=\gamma_{E|0\rangle}$. Note that whilst ISC selection rules likely imply
that the populations rates could differ, such differences will not significantly effect the application of the rate
model in the present section.

\begin{figure}
  \centering
  \includegraphics{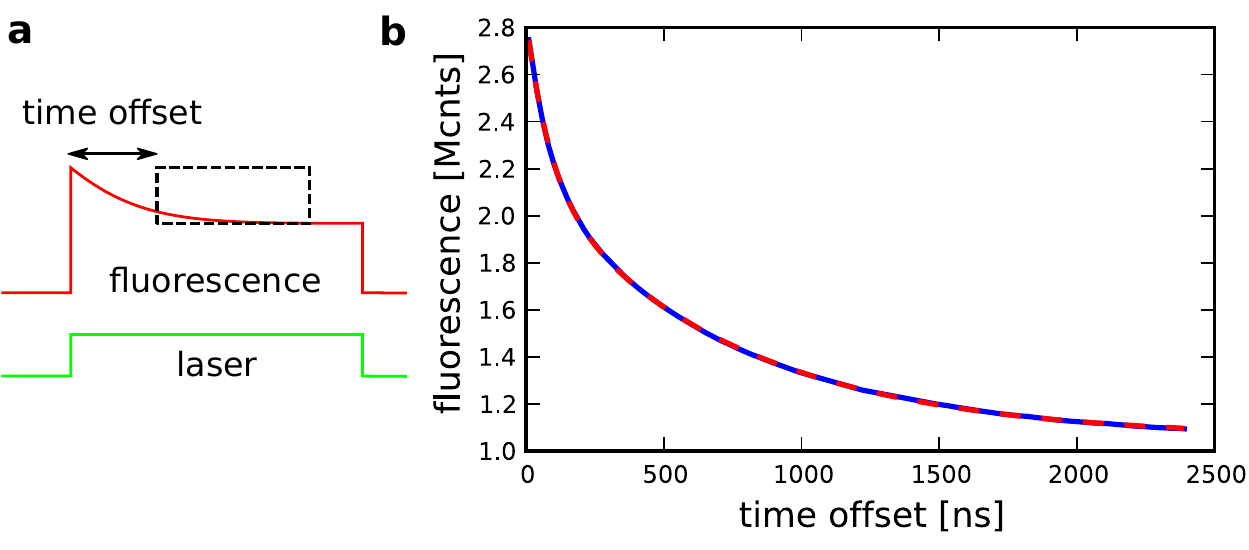}
  \caption{\label{fig:triplet_population}\textbf{Determination of the overall triplet population rate by pulsed
optical excitation.} a) Schematic of the measurement and data analysis. The
stimulation consits of optical pulses that are separated by idle times. Both
are longer than the triplet lifetimes. The resulting fluorescence response
profile is processed by integrating the fluorescence over a window of fixed
length and moving this window over the profile, starting from the initial
rising edge. b) Experimental data (blue solid line) and double exponential fit
(red dashed line), providing two time constants $\tau_1=85$ns and
$\tau_2=724$ns.}
\end{figure}
A second method for determining the triplet population rates is to use a pulsed optical experiment.
We expect that the population rates of the
triplet sublevels are similar for all three levels and the time scales of population and decay are well seperated
(the population rates and decay rates differ by about an order of magnitude).
If this was not the case, we would expect a more complicated decay behavior in the
lifetime measurements. In this case, the total transition rate from
the excited state to the triplet levels
can be determined from the fluorescence response of the system to a single optical pulse as shown
in Fig.\ref{fig:triplet_population}a. The length of the laser pulses and the idle time between two pulses is
much longer than the triplet lifetime. If the laser is set to high
power such that the optical transitions is saturated, the ground state is essentially
depleted and first all population resides in the excited state. The population then starts to
leak to the triplet state. Thus, the length
of the excitation pulse determines how much
population is transferred. Since the lifetimes of the triplet state are much longer than the population rates, recycling
of population through the triplet and ground state during this process is negligable.
Therefore, an approximate exponential increase of the
triplet population is expected with time constant $\tau_{ET}=\gamma_{ET}^{-1}$. As before, the triplet
population can be measured by a readout laser pulse and recording the fluorescence
response profile. In the present measurement, the initialization and readout laser pulses
can be merged to a single pulse and the distinction between the two pulses is achieved by moving
the integration window for the determination of the triplet population over the pulse profile (refer to
figure~\ref{fig:triplet_population}a).
The result of this measurement is shown in Figure~\ref{fig:triplet_population}b.
A double exponential decay is observed. The short time constant reflects the buildup of the triplet population.
The longer time constant reflects the time required to reach the steady state under optical pumping
and is given by a combination of the different decay rates. The combined population rate
is found from the short time constant as $\gamma_{ET}\approx(85\,\textrm{ns})^{-1}$. This value agrees reasonably well
with the value of about $56$~ns obtained from the photon auto-correlation. Appreciating both independent methods, we
estimate $\gamma_{ET}\approx(70(20)\,\textrm{ns})^{-1}$.

\subsection{Application of the rate equation model}

In this section we provide further detail about the rate equation model. In particular, we present that the recovery of the ground state population as observed in Fig 3 in the main text is the evidence for the triplet meta stable state and analyze how the observed
amplitudes of exponentially decaying
components of the triplet levels are related to the rate parameters. To this end, the analytical solutions for the
ground state spin recovery without and with $\pi$--pulses will be introduced.
The steady state solutions based on the rate model can be easily found (not shown)
and can be considered as initial values at $t=0$ for the ground state recovery measurement. Then the spin populations of
each states after time $\tau'$ (while $\gamma_{GE}=0$) can be found. In the experiments, a 6 $µ$s long readout pulse
was applied at $t=\tau'$ and the fluorescence response was integrated only over a short ($20\sim 200$~ns) period at
the very beginning of the readout window. Because no significant difference in the integrated signal was found up to a
200 ns long integration window, we suppose that the repolarization hardly affected so that the integrated intensity is
directly proportional to the ground state spin population after $\tau '$. Thus the ground state spin population at
$t=\tau '$ is only of our interest, which is
\begin{eqnarray}\label{eq:pop_g}
n_{G}(\tau ') = C_{1} \cdot n_{G}(0) \cdot e^{-\Gamma \cdot \tau '} - \sum_{i=|+\rangle,|-\rangle,|0\rangle}
\frac{\Gamma}{\Gamma-\gamma_{iG}} \cdot n_{i}(0) \cdot e^{-\gamma_{iG} \cdot \tau '} + C_{0} \cdot n_{G}(0)
\end{eqnarray}
where $n_{i}(0)$ is the initial value for each state which is equivalent to the steady state value,
$\Gamma=\gamma_{EG}+\gamma_{E|+\rangle}+\gamma_{E|-\rangle}+\gamma_{E|0\rangle}$, $C_{1}=-\frac{\gamma_{EG}\cdot
\gamma_{GE}}{\Gamma^{2}} +\sum_{i=|+\rangle,|-\rangle,|0\rangle} \frac{\gamma_{iG}\cdot \gamma_{Ei}\cdot
\gamma_{GE}}{\Gamma^{2} \cdot (\Gamma-\gamma_{iG})}$, and $C_{0}=1-C_{1} +\sum_{i=|+\rangle,|-\rangle,|0\rangle}
\frac{\gamma_{Ei}\cdot \gamma_{GE}}{\gamma_{iG} \cdot (\gamma_{EG}-\gamma_{iG})}$.
When an ideal resonant $\pi$--pulse between two sublevels, $|+\rangle$ and
$|-\rangle$ for instance, is applied right after the initialization pulse,
the initial population of these two states are exchanged. The solution for
this case becomes
\begin{eqnarray}\label{eq:pop_g_mw}
n_{G,|+\rangle|-\rangle}(\tau ') &=& C_{1} \cdot n_{G}(0) \cdot e^{-\Gamma \cdot
\tau '} - \sum_{i,j=|+\rangle,|-\rangle,i\neq j} (n_{j}(0)+
\frac{\gamma_{iG}}{\Gamma-\gamma_{iG}} \cdot n_{i}(0) ) \cdot
e^{-\gamma_{iG} \cdot \tau '} \nonumber \\
&&-\frac{\Gamma}{\Gamma-\gamma_{|0\rangle G}} \cdot n_{|0\rangle}(0) \cdot
e^{-\gamma_{|0\rangle G} \cdot \tau '}  + C_{0,|+\rangle|-\rangle} \cdot
n_{G}(0)
\end{eqnarray}
where $C_{0,|+\rangle|-\rangle}$ is a constant similar with $C_{0}$. Solutions for the two other
cases $n_{G,|+\rangle|0\rangle}(\tau ')$ and $n_{G,|-\rangle|0\rangle}(\tau ')$
also can be found (not shown). Both solutions consist of four exponential terms,
but the first term has little contribution because its decay is faster than the
others ($\Gamma>\gamma_{EG}\gg\gamma_{iG}$) by at least one order of magnitude,
and its amplitude is very small too as will be seen later. In addition, because
the found triplet state life times are at least one order of magnitude slower
than the excited singlet state life time, we get
$\gamma_{iG}/(\Gamma-\gamma_{iG}) \ll 1$ and $\Gamma/(\Gamma-\gamma_{iG}) \approx
1$. Thus the above solutions can be rewritten as
\begin{eqnarray}
n_{G}(\tau ') \approx -\sum_{i=|+\rangle,|-\rangle,|0\rangle} n_{i}(0) \cdot
e^{-\gamma_{iG} \cdot \tau
'}+C_{0} \cdot n_{G}(0)  ,
\end{eqnarray}

\begin{eqnarray}
n_{G,|+\rangle|-\rangle}(\tau ') \approx -\sum_{i,j=|+\rangle,|-\rangle,i\neq j}
n_{j}(0) \cdot e^{-\gamma_{iG}
\cdot \tau '}-n_{|0\rangle}(0) \cdot e^{-\gamma_{|0\rangle G} \cdot \tau
'}+C_{0,|+\rangle|-\rangle} \cdot n_{G}(0).
\end{eqnarray}

Thus one can conclude that the observed ground state recovery curve can be explained well by three exponential decay
terms of which decay constants are the decay constants of the triplet sublevels or even the double exponential decay if
one of the three decay terms is too small. And the inserted $\pi$--pulse does not alter the decay constants but only
exchanges the amplitudes of the sublevels under the influence of the applied $\pi$--pulse as long as the excited singlet
state life time is faster than the triplet state life times. Therefore, one can assign the measured decay constants to
each sublevels by comparing the amplitudes of each decay terms obtained with and without applying $\pi$--pulses. In
order to make an indisputable conclusion, it is better to compare the numerically simulated amplitudes of each decaying
components with the observed ones.

\begin{figure}
\centering
\includegraphics[width=1\columnwidth]{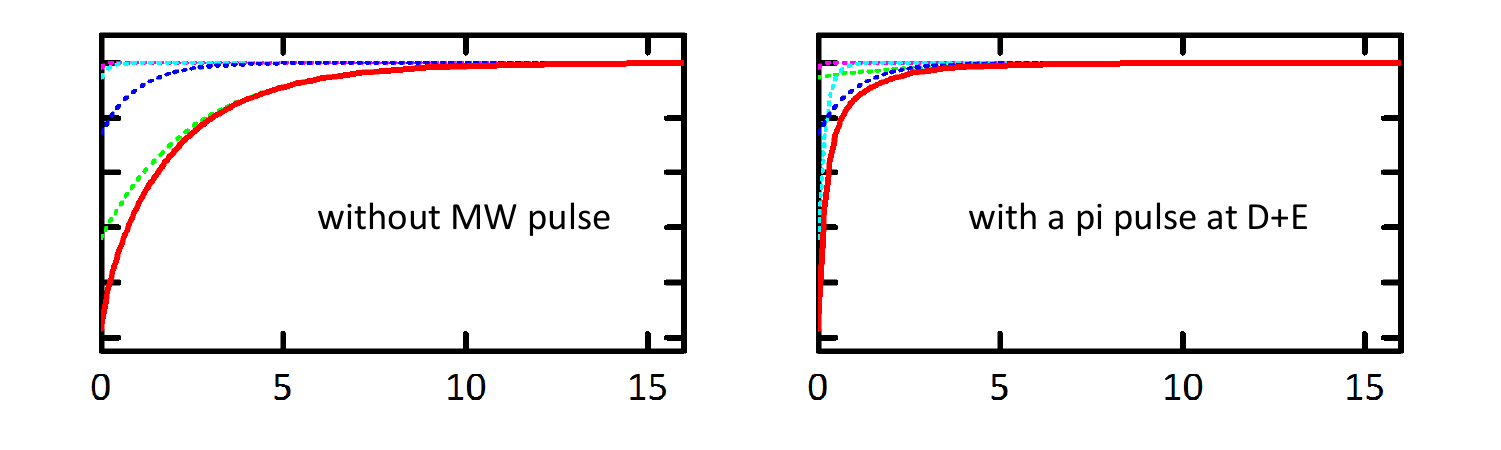}
\caption{\label{fig:T1_simul}\textbf{Numerical simulations of the ground state recovery measurements.} Red lines are the
simulated recovery curves, cyan, blue, and green curves are the exponential decays of $|+\rangle$, $|-\rangle$, and
$|0\rangle$, respectively. The first exponential terms (violet lines) in equation~(\ref{eq:pop_g}) and
(\ref{eq:pop_g_mw}) are almost ignorable because of their fast decay and small amplitudes.}
\end{figure}

The numerical simulations using the exact solution, equations~(\ref{eq:pop_g})
and (\ref{eq:pop_g_mw}) with the set of
parameters of $\gamma_{EG}=(10\,\textrm{ns})^{-1}$, $\gamma_{GE}=(10\,\textrm{ns})^{-1}$,
$\gamma_{E|+\rangle}=\gamma_{E|-\rangle}=\gamma_{E|0\rangle}=(170\,\textrm{ns})^{-1}$, $\gamma_{|+\rangle
G}=(200\,\textrm{ns})^{-1}$,
$\gamma_{|-\rangle G}=(1000\,\textrm{ns})^{-1}$, $\gamma_{|0\rangle G}=(2500\,\textrm{ns})^{-1}$ are shown in
figure~\ref{fig:T1_simul}. Note that the common value of $170\,\textrm{ns}$ was used for three population rates and the
typical value for the optical pumping rate $\gamma_{GE}=(10\,\textrm{ns})^{-1}$ was used. The first term in
equations~(\ref{eq:pop_g}) and (\ref{eq:pop_g_mw})
(violet curves in figure~\ref{fig:T1_simul}) of which decay constant is
$\Gamma=\gamma_{EG}+\gamma_{E|+\rangle}+\gamma_{E|-\rangle}+\gamma_{E|0\rangle}$, is negligible because it decays very
quickly and the amplitude is minute. Among the remaining three exponential decays originated from the decays of the
triplet sublevels, the amplitude of the fastest decay is smaller than others, so this decay is hardly detectable in
measured data as already shown above. Overall, these simulated curves are in
close agreement with the measured ones and can explain the measured curves
very well.
The simulated and measured amplitudes of each decay terms are summarized in table~\ref{tab:comparison}. The measured
amplitudes without applying $\pi$--pulses are in very good agreement with the
simulations. The exchange of the initial
populations by the coherent $\pi$--pulses are well described in both measured
and simulated values, and they are in good
agreement. The found discrepancy between the size of the simulated and the measured amplitudes can have a few origins:
intersystem crossing among each sublevels, non-ideal $\pi$--pulses.
\begin{table}[htbp]
  \centering
    \begin{tabular*}{0.75\textwidth}{@{\extracolsep{\fill}}ccccccc}
    \toprule
          & \multicolumn{3}{c}{simulated} & \multicolumn{3}{c}{measured} \\
    \midrule
          & A+    & A-    & A0    & A+    & A-    & A0 \\
		\midrule
    without MW & 0.2   & 1.0   & 2.4   & -     & 1.0(1) & 2.3(1) \\
    D+E & 2.4   & 1.0   & 0.2   & 1.11(4) & 1.00(3) & - \\
    D-E & 0.2   & 2.4   & 1.0   & -     & 1.7(1) & 1.0(1) \\
    2E & 1.0   & 0.2   & 2.4   & 1.00(9) & -     & 10.16(4) \\
    \bottomrule
		\end{tabular*}%		
	\caption{\textbf{Comparison of the measured amplitudes with the simulations.} The observed switching of the
initial populations by $\pi$--pulses are in consistent with the simulations.}
  \label{tab:comparison}%
\end{table}%

\subsection{Conclusions for the electronic structure}

As discussed in the previous sections, the observed behaviour of the optical fluorescence implies that the optically
detected
magnetic resonance occurs within a metastable spin triplet level that exists energetically between the ground and
optically excited levels of the defect. It was also discussed in the main text that the nuclear resonance of the
$^{13}C$ nuclear spin
performed whilst the defect is in the electronic ground state implies that the ground electronic level is a spin
singlet. These facts can be used to conclude various aspects of the defect's electronic structure.

\begin{figure}
\centering\includegraphics[width=0.6\columnwidth]{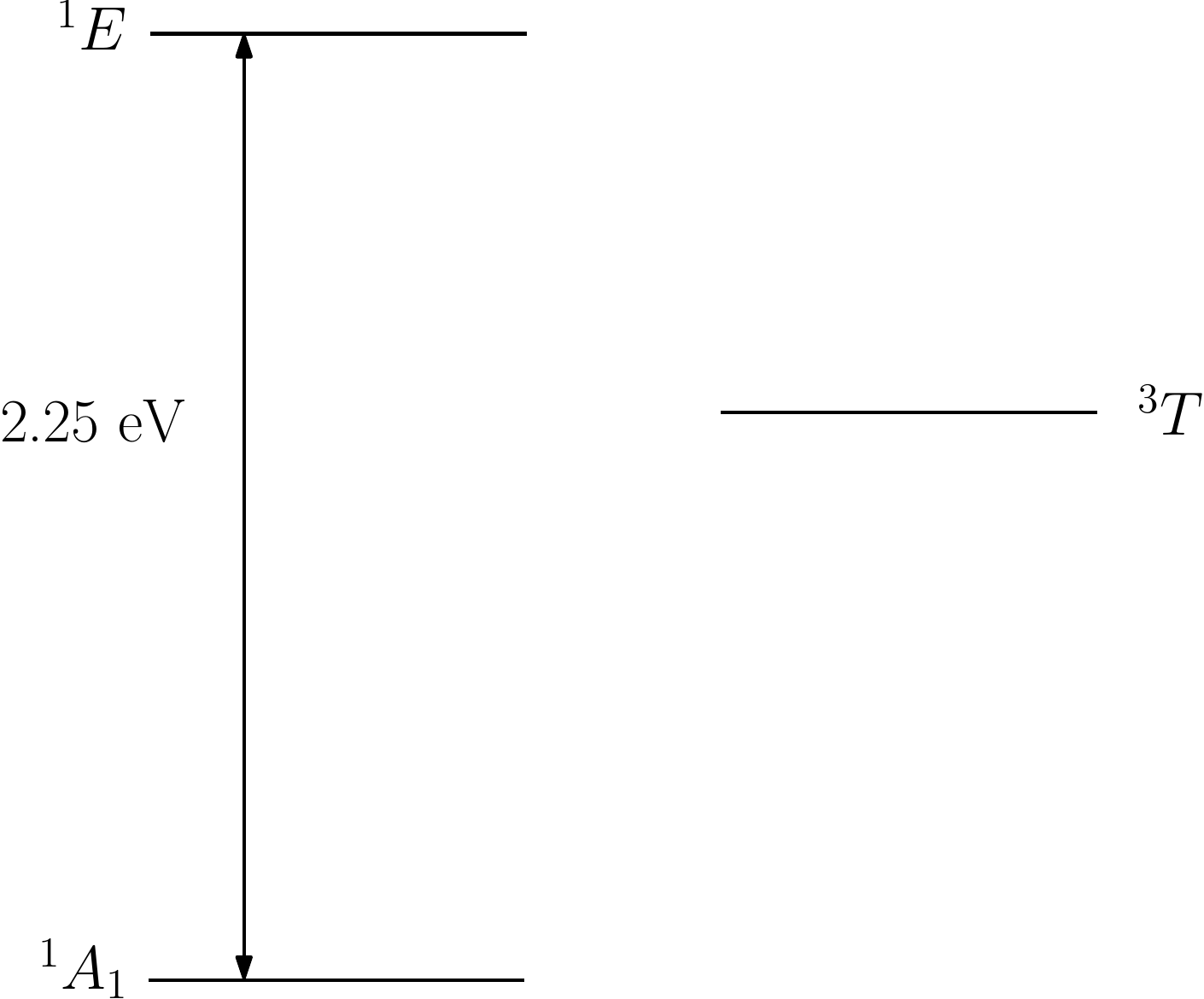}
\caption{\label{fig:electronicstructure}\textbf{Observed electronic structure of the ST1 defect}. Levels are labelled by
their appropriate term symbols. Note that the orbital symmetries of $^1E$ and $^3T$ are unknown. The vertical arrow
denotes the optical ZPL at 550 nm (2.25 eV). The estimate $\sim0.15$ eV obtained for the lattice relaxation energy in
the previous section can be combined with the ZPL to estimate an energy of $\sim2.40$ eV between the $^1A_1$ and $^1E$
electronic levels. Other electronic levels may be present, but are yet to be detected.}
\end{figure}

According to Hund's rules, a spin singlet ground state can only occur if the ground electronic configuration is closed
(i.e. filled electron orbitals) \cite{tinkham1964group}. Using the electron orbital notation of figure
\ref{fig:bandstructure},
the ground electronic configuration is $a^2$, where all lower energy orbitals (i.e. orbitals of the valence band) are
taken to be filled. Applying Unsold's theorem, a closed configuration is necessary orbitally symmetric (denoted $A_1$)
\cite{tinkham1964group}, such that the term symbol of the ground electronic level is $^1A_1$. Optical excitation
promotes one of
the electrons to the higher energy orbital $b$, such that the configuration of the optically excited level is $ab$. Spin
selection rules of optical transitions imply that the optically excited state is also a spin singlet. As the symmetry of
the orbitals are unknown, it is not possible to determine the orbital symmetry of the optically excited state, so for
convenience it will be denoted by the term symbol $^1E$. The spin
triplet level may be of the same configuration as $^1E$ (the triplet being lower in energy in accordance with Hund's
rules) or another excited configuration. Consequently, the orbital symmetry of the triplet level is also unknown and it
will be labelled by the convenient term symbol $^3T$. There may also exist other electronic levels between the triplet
and singlet levels and also above $^1E$ (which are relevant as they may be populated under 532 nm excitation). However,
such additional levels have not been detected thus far. The aspects of the electronic structure that can be concluded
from our observations are depicted in figure \ref{fig:electronicstructure}. Future optical polarization and stress
studies can be used to determine the orbital symmetry of the optically excited singlet level \cite{Davies1976}.

\begin{figure}
\centering\includegraphics[width=0.6\columnwidth]{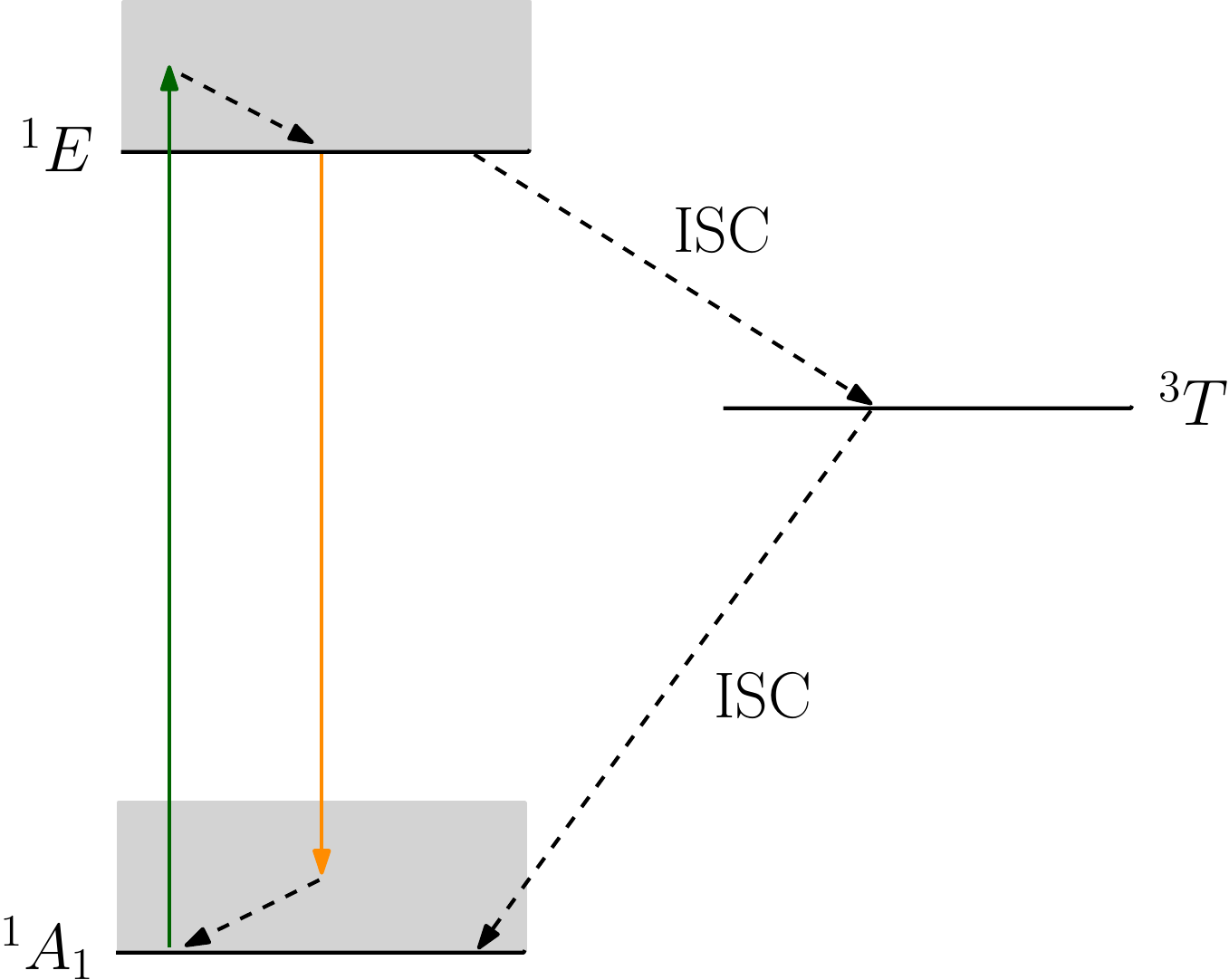}
\caption{\label{fig:opticalcycle}\textbf{Model of the optical cycle of the ST1 defect}. Known electronic levels are
denoted by solid lines. The continuum of vibronic levels that give rise to the optical absorption and emission bands are
depicted as shaded regions. Solid vertical arrows denote radiative transitions. Dashed arrows denote non-radiative decay
within the vibronic levels of $^1A_1$ and $^1E$ as well as the ISCs between the singlet and triplet levels.}
\end{figure}

Given the observed aspects of the defect's electronic structure, a working model can be established of the defect's
optical cycle (refer to figure \ref{fig:opticalcycle}). Beginning in the ground state, optical excitation transfers the
defect into a vibronic level of $^1E$, which is followed by rapid non-radiative decay through the $^1E$ vibronic levels,
such that thermal equilibrium is established within the level's lifetime. From the $^1E$ level, the defect may decay
radiatively back to the ground state or non-radiatively via an intersystem crossing (ISC) to the metastable triplet
level. The defect decays from the $^3T$ level by a subsequent ISC to the ground state. The ISCs are allowed via
spin-orbit coupling of the triplet and singlet levels and are mediated by phonon emission \cite{Loubser1978}. The nature
of spin-orbit coupling implies that the ISCs will be to some degree dependent on triplet spin-projection. The net effect
being different storage times in the triplet level for different spin-projections
that results in a modulation of the optical fluorescence if the defect is transferred between the spin-projections of
the triplet. The presence of other electronic levels in the optical cycle will to some degree change the behavior of the
cycle from that described by the model depicted in figure \ref{fig:opticalcycle}. No such variance has been detected to
date.

\section{Triplet level spin-Hamiltonian, zero-field splittings and spin axes}
\label{sec:spinhamiltonian}

The spin-Hamiltonian that is appropriate to model the fine structure of the triplet level takes the general
form~\cite{Loubser1978,stoneham2001theory}
\begin{equation}
\hat{H} = \vec{S}\cdot\mathbf{D}\cdot\vec{S}+\vec{S}\cdot\mathbf{g}\cdot\vec{B}
\end{equation}
where $\vec{S}$ is the dimensionless $S=1$ spin operator, $\vec{B}$ is the magnetic field at the defect, and
$\mathbf{D}$ and $\mathbf{g}$ are each $3\times3$ tensors to be determined experimentally. The first term describes the
effects of the crystal field that typically arise from electronic spin-spin interactions and/ or second-order spin-orbit
interactions. The second term describes the electronic Zeeman interaction as governed by the tensor $\mathbf{g}$, which
typically differs from the isotropic g-factor of a free-electron $g_e=2.0023$ by orbital angular momentum and spin-orbit
effects.

\begin{figure}
  \centering
  \includegraphics[width=0.6\columnwidth]{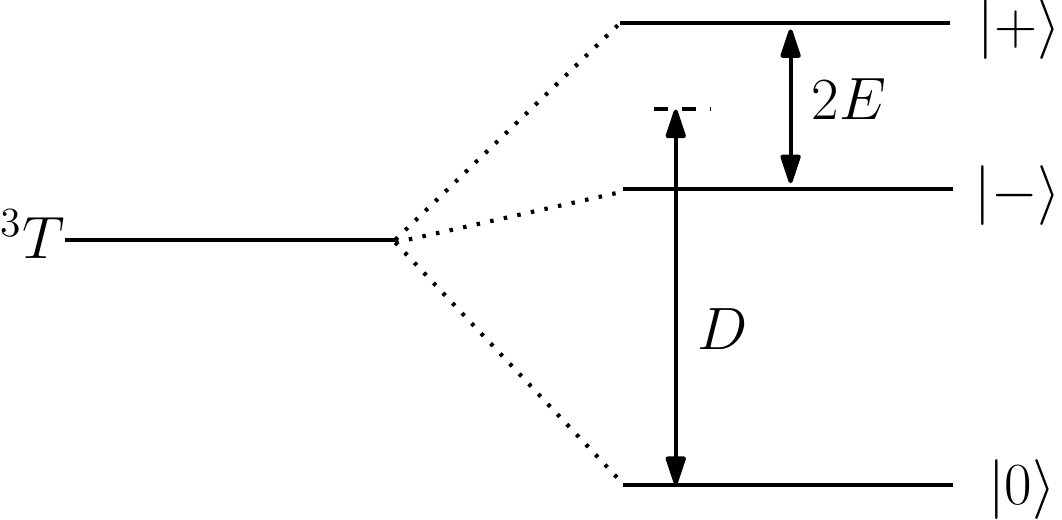}
  \caption{\textbf{Triplet level zero field fine structure.} Note that $D$ and $E$ have been assumed positive. In terms
of the spin states $|m_s\rangle$ with triplet spin-projection $m_s$, the zero field eigenstates are $|+\rangle =
(|+1\rangle+|-1\rangle)/\sqrt{2}$, $|-\rangle = (|+1\rangle-|-1\rangle)/\sqrt{2}$ and $|0\rangle$.}
  \label{fig:zffinestructure}
\end{figure}

\begin{figure}
\centering\includegraphics[width=1\columnwidth]{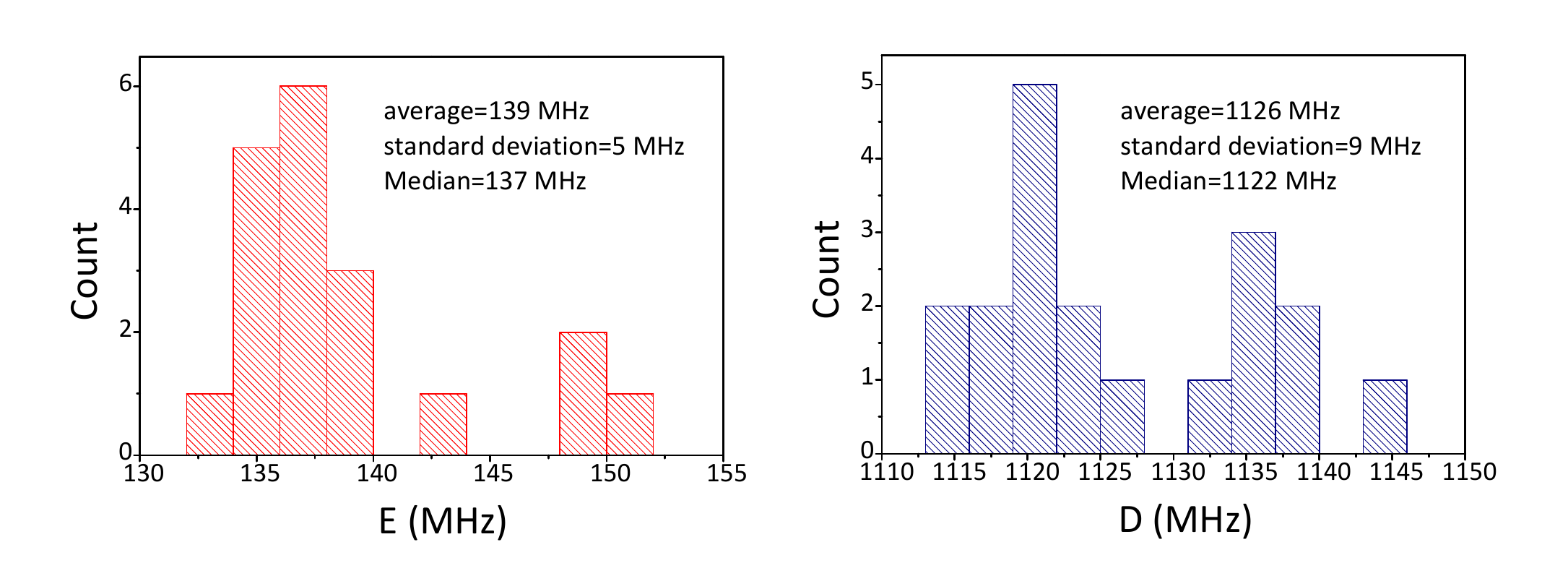}
\caption{\label{fig:statistics}\textbf{Distributions of the ZFS of 19 ST1 defects.} The variance of the parameters is
consistent with a distribution of strain within the nano-wires.}
\end{figure}

By an appropriate choice of the defect coordinate system $(xyz)$, the crystal field tensor $\mathbf{D}$ may be
diagonalized.
Furthermore, as our experiments did not detect any anisotropy in the electronic Zeeman interaction, the tensor
$\mathbf{g}$ may be reduced to an isotropic g-factor. The simplified spin-Hamiltonian is
thus~\cite{Loubser1978,stoneham2001theory}
\begin{equation}
\hat{H} = D[S_z^2-S(S+1)/3]+E(S_x^2-S_y^2)+g\vec{S}\cdot\vec{B}
\label{eq:spinhamiltonian}
\end{equation}
where $D$ and $E$ are zero field splitting (ZFS) parameters and $g=2.0(1)$ is the observed isotropic g-factor. The zero
field fine structure of the triplet level that is described by the above spin-Hamiltonian is depicted in figure
\ref{fig:zffinestructure}. The distributions of the observed ZFS parameters of 19 defects are shown in
figure~\ref{fig:statistics}. The relatively small variance of the parameters strongly indicates that they are intrinsic
properties of the ST1 defect that are slightly modified by the local strain of each defect site. Note that the
imprecision of the observed g-factor is due to the experimental uncertainty of the magnetic field calibration.
Also note that a precise measurement of the g-factor could give additional insight into the symmetry of the defect:
Anisotropy of the g-factor arises due to spin-orbit coupling between the electronic levels and these couplings obey
symmetry selection rules. C2v symmetry implies negligible spin-orbit coupling and thus likely an isotropic
g-factor, while in case of a symmetry lower than C2v, an anisotropic g-factor could be present. The
precise sign of the ZFS parameters can not be determined from our measurements.

\begin{figure}
  \centering
  \begin{subfigure}[c]{0.4\textwidth}
    \textbf{a}

    \includegraphics[width=0.8\columnwidth]{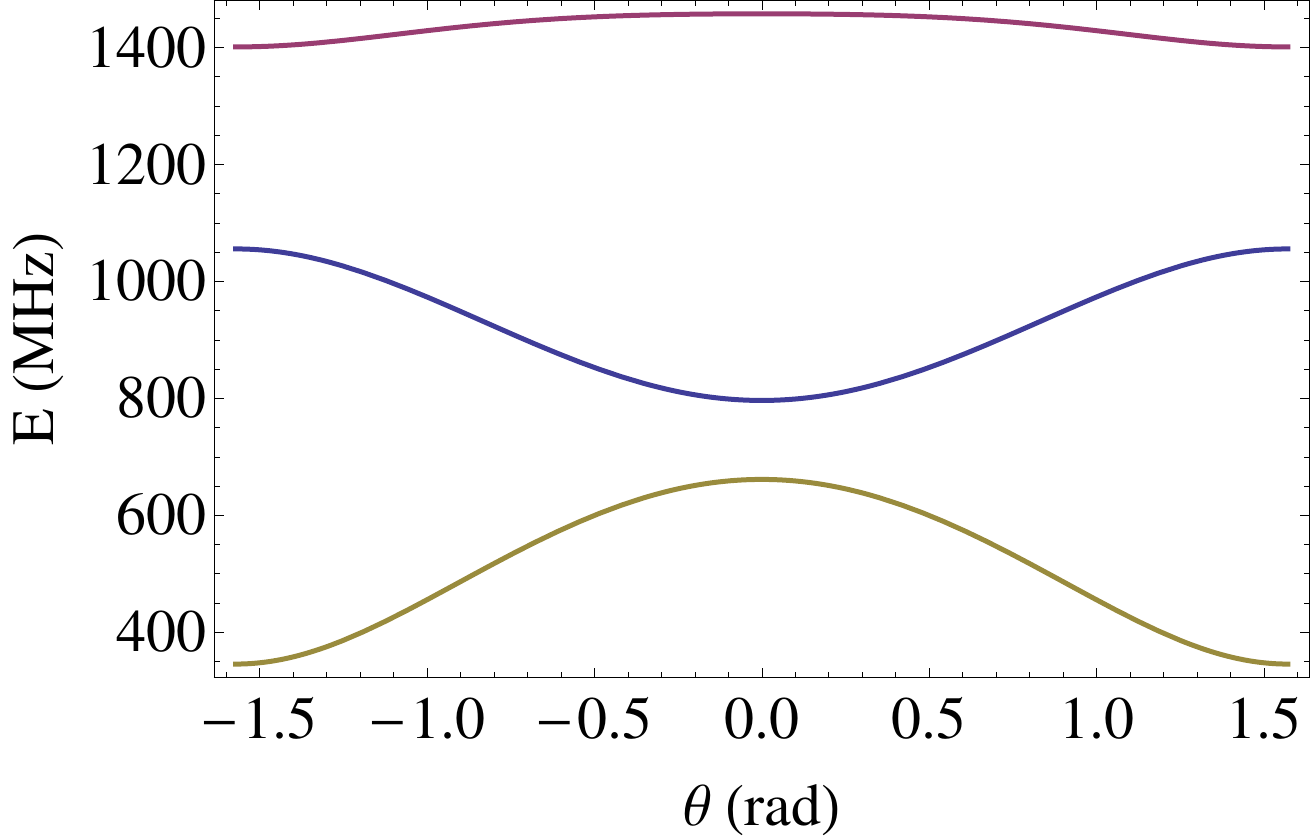}

    \textbf{b}

    \includegraphics[width=0.8\columnwidth]{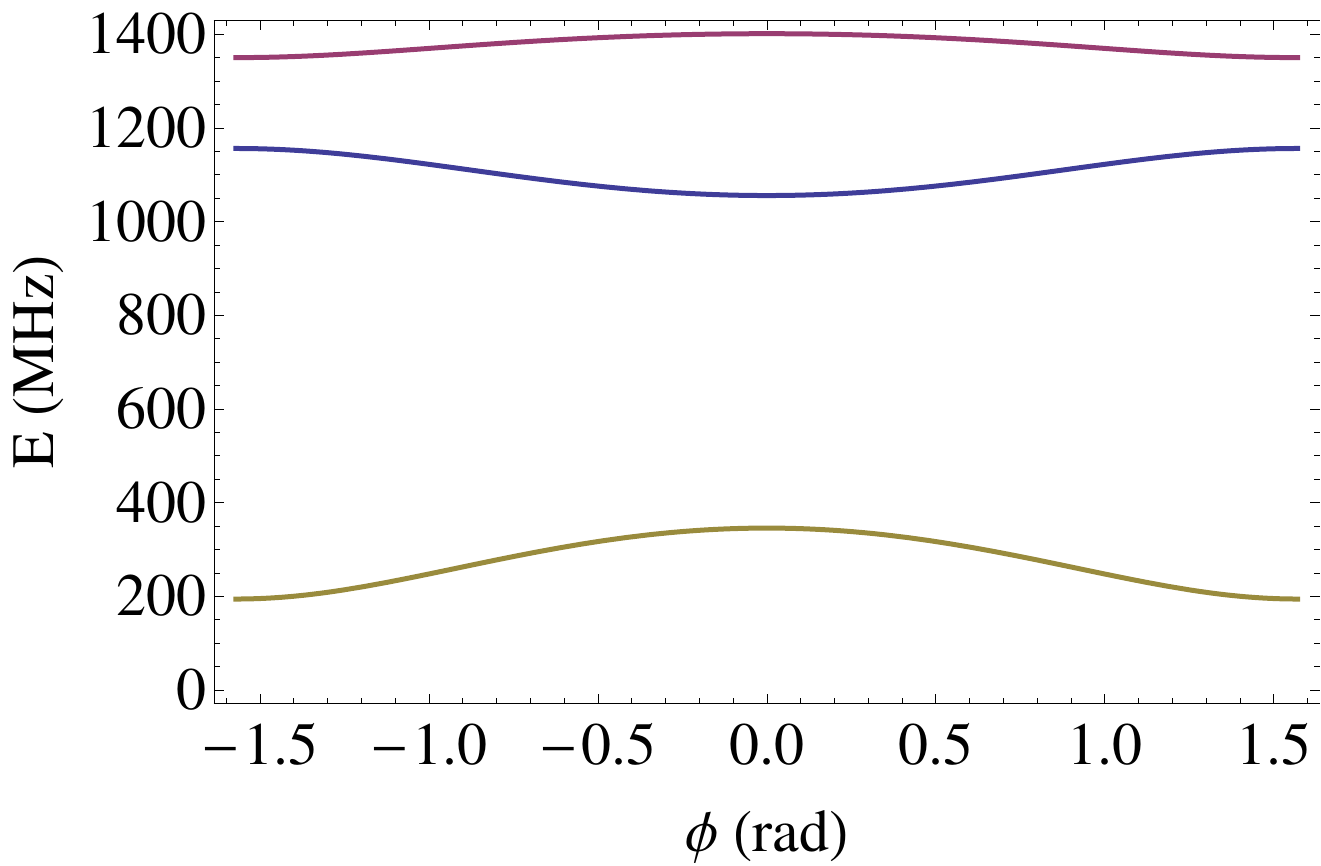}
  \end{subfigure}
  \begin{subfigure}[c]{0.4\textwidth}
    \textbf{c}

    \rule{1em}{0mm}\includegraphics[width=0.9\columnwidth]{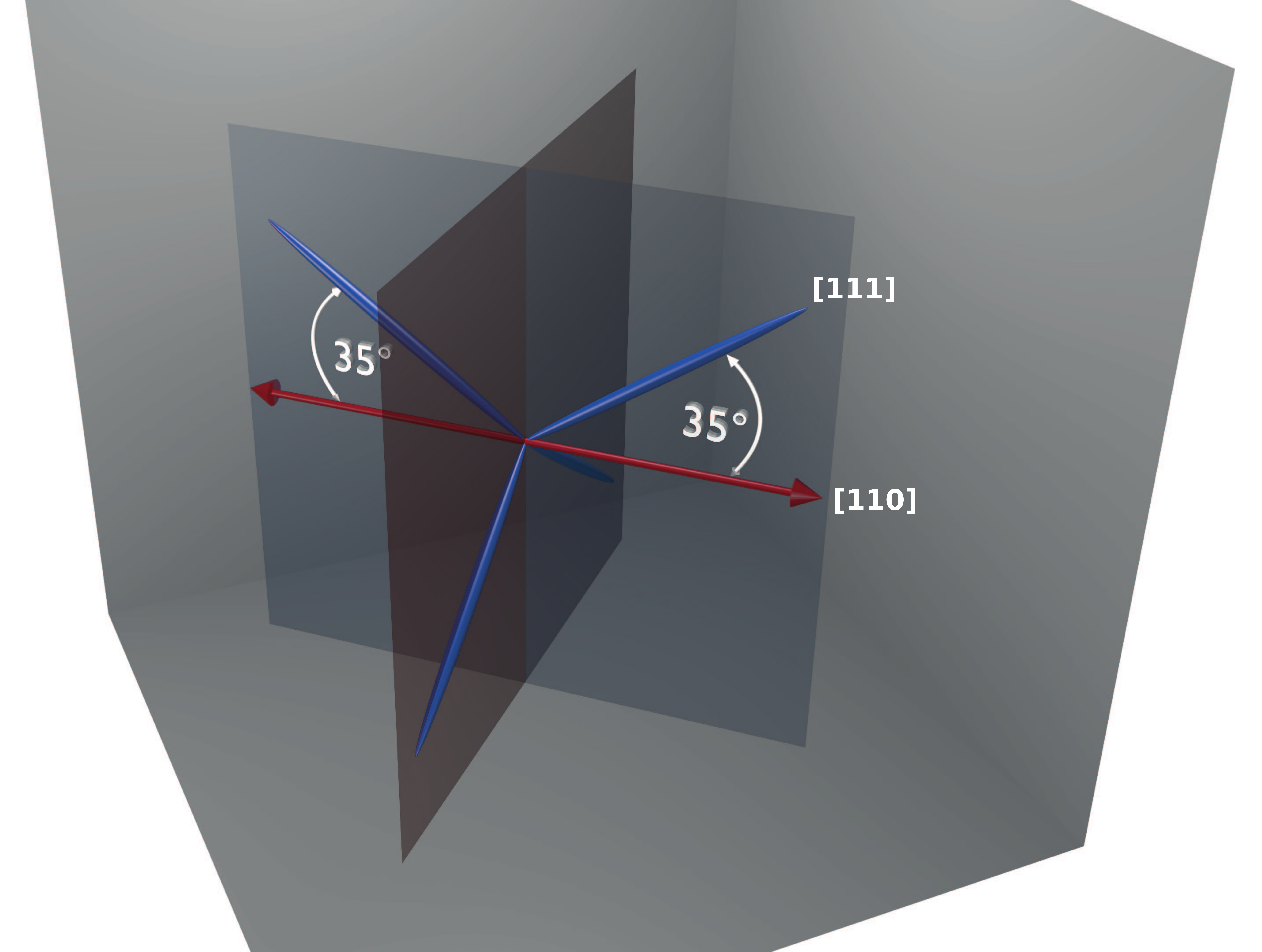}
  \end{subfigure}
  \caption{\textbf{Magnetic field alignment and major symmetry axis.} The plots depict the variation of the three spin
resonances as a function of magnetic field orientation: (a) $B_x=B\sin\theta$, $B_y=0$, $B_z=B\cos\theta$; and (b)
$B_x=B\cos\phi$, $B_y=B\sin\phi$, $B_z=0$. The values $D=1128$ MHz, $E=139$ MHz and $B=300$ MHz were used to produce the
plots. Relationships between the resonances are evident when the magnetic field is aligned with the defect's $z$ axis
($\theta=0$), $x$ axis ($\phi=0$) and $y$ axis ($\phi=\pi/2$). (c) Major symmetry axis of the ST1 defect. Blue
lines indicate the four NV axes in the diamond crystal.}
  \label{fig:fieldalignment}
\end{figure}

The spin transition frequencies for different orientations of a static magnetic field are depicted in figure
\ref{fig:fieldalignment}. The spin resonances have clear relationships when the magnetic field is aligned along one of
the defect's coordinate axes. The orientation of the magnetic field can be simultaneously measured with respect to the
crystal coordinate system using three inequivalent nearby NV$^-$ centers and knowledge of their [111] symmetry axis
(method described elsewhere~\cite{Steinert2010}). Consequently, the ST1 defect's major ($z$) axis was determined by
monitoring the ODMR frequencies of a chosen ST1 defect and nearby NV$^-$ centers whilst manipulating the orientation of
the magnetic field generated by an electromagnet placed above the sample. The elicited major axis together with the four
NV$^{-}$ orientations is shown in figure~\ref{fig:fieldalignment}c. The major axis is on a plane defined by two
NV$^{-}$
axes (NV1 and NV2 axes in this plot) and off by $35(1)\,^{\circ}$ with respect to
these two axes and perpendicular to the other two NV$^{-}$ axes. These findings lead to the conclusion that the major
axis is approximately in the [110] crystal direction. The same conclusion has been obtained from several other ST1
defects.

The ZFS parameters of molecule-like defects in diamond, such as the NV$^-$ center, are predominately due to spin-spin
interactions, whereas the ZFS parameters of atom-like defects, such as transition metal ion impurities, arise largely
from spin-orbit interactions~\cite{Loubser1978,stoneham2001theory}. Assuming that the ZFS parameters arise from
spin-spin
interaction between the two unpaired electrons of the triplet level, the ZFS parameters provide insight into the
distribution of unpaired spin density associated with the triplet level via the following
expressions~\cite{stoneham2001theory}
\begin{eqnarray}
D = \frac{3}{2}g_e^2µ_B^2\frac{µ_0}{4\pi}\langle\frac{1-3(z_{12}^2/r_{12}^2)}{r_{12}^3}\rangle \nonumber \\
E = -\frac{3}{2}g_e^2µ_B^2\frac{µ_0}{4\pi}\langle\frac{x_{12}^2-y_{12}^2}{r_{12}^5}\rangle
\end{eqnarray}
where $µ_0$ is the vacuum permeability, $µ_B$ is the Bohr magneton, $x_{12}$, $y_{12}$, $z_{12}$ are the coordinates
of the displacement vector $\vec{r}_{12}$ with magnitude $r_{12}$ that connects the positions of the two electrons, and
the angle brackets denote the direct integral over the unpaired spin density. Note that the exchange integral is
neglected in the above.

The interpretation of the observed ZFS parameters of the ST1 defect using the above expressions indicates that the
unpaired spin density is elongated along the major ($z$) axis of the defect and is not symmetrical in the $x-y$ plane of
the defect ~\cite{stoneham2001theory}. To explain this interpretation, an analogy may be drawn with the NV$^-$ center,
where the
ground state unpaired spin density is equally distributed over the three dangling sp$^3$ bonds of the carbon atoms
surrounding the vacancy~\cite{Loubser1978}. The trigonal symmetry of the NV$^-$ center implies that its ground state
spin
density is symmetrical in the (111) plane in the absence of strain, but is elongated along the [111] direction. The $E$
parameter of NV$^-$ is therefore zero in the absence of strain, whilst the $D$ parameter is non-zero. In the presence of
strain, the trigonal symmetry of the NV$^-$ center is lowered and its ground state spin density is distorted in the
(111) plane, thereby resulting in a non-zero $E$ parameter. The non-zero
$E$ parameter of the ST1 defect implies that it is a defect of less than trigonal symmetry
~\cite{Loubser1978,stoneham2001theory}.
Combining this conclusion with the orientation of the defect's major axis along the [110] direction restricts the set of
possible point group symmetries of the ST1 defect to just including $C_{2v}$ and its sub-groups: $C_2$, $C_{1h}$, and
$C_1$.

The value of $D$ for the ST1 defect may also be used to crudely estimate an upper bound of 4 angstroms for the average
separation $\langle r_{12}\rangle$ of the two unpaired electrons~\cite{Loubser1978,stoneham2001theory}. A similar
estimate applied
to the ground state of the NV$^-$ center yields 3 angstroms~\cite{He1993}. By comparison, it can be concluded that the
spin density of the ST1 defect is likely to be distributed over a greater region than the ground state spin density of
the NV$^-$ center. The ZFS parameters are thus consistent with an unpaired spin density distributed over the orbitals of
several atomic sites along the [110] crystal direction. The ZFS parameters are thus consistent with some degree of
preferential distribution of the unpaired spin density over the orbitals of several atomic sites along the [110] crystal
direction.

\section{Triplet hyperfine structure}
\label{sec:hyperfinestructure}

\begin{figure}
  \centering
  \includegraphics[width=0.9\columnwidth]{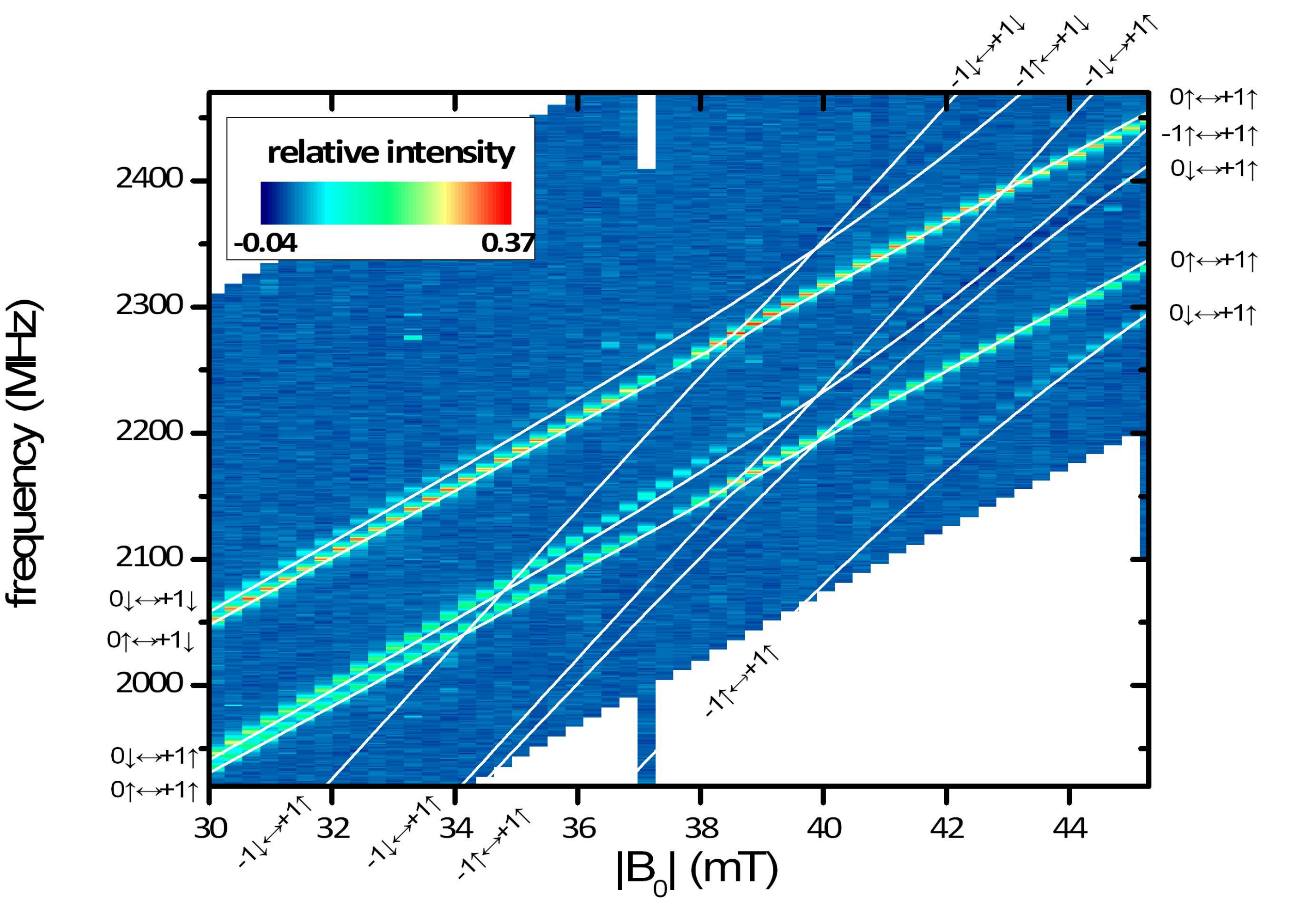}
  \includegraphics[width=0.9\columnwidth]{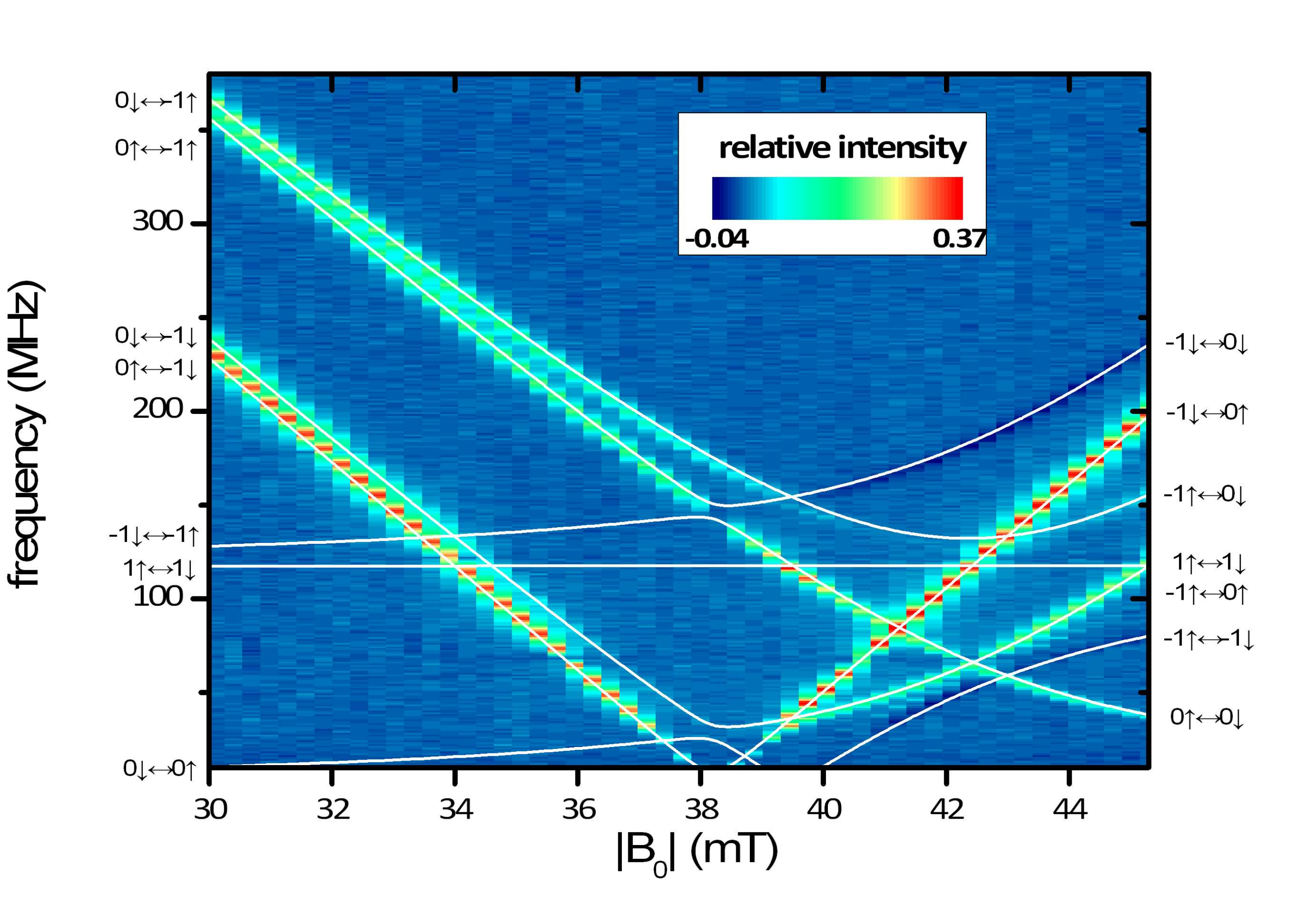}
  \caption{\textbf{Hyperfine structure at the LAC.} Upper panel: $D+E$ transition, lower panel: $D-E$ transition. The
color maps show experimental ODMR data. White lines show theoretical results. The approximate eigenstates linking the
transitions are denoted on the sides of the plots.}
\label{fig:hyperfine_experiment}
\end{figure}

\begin{figure}
  \centering
  \includegraphics[width=0.9\columnwidth]{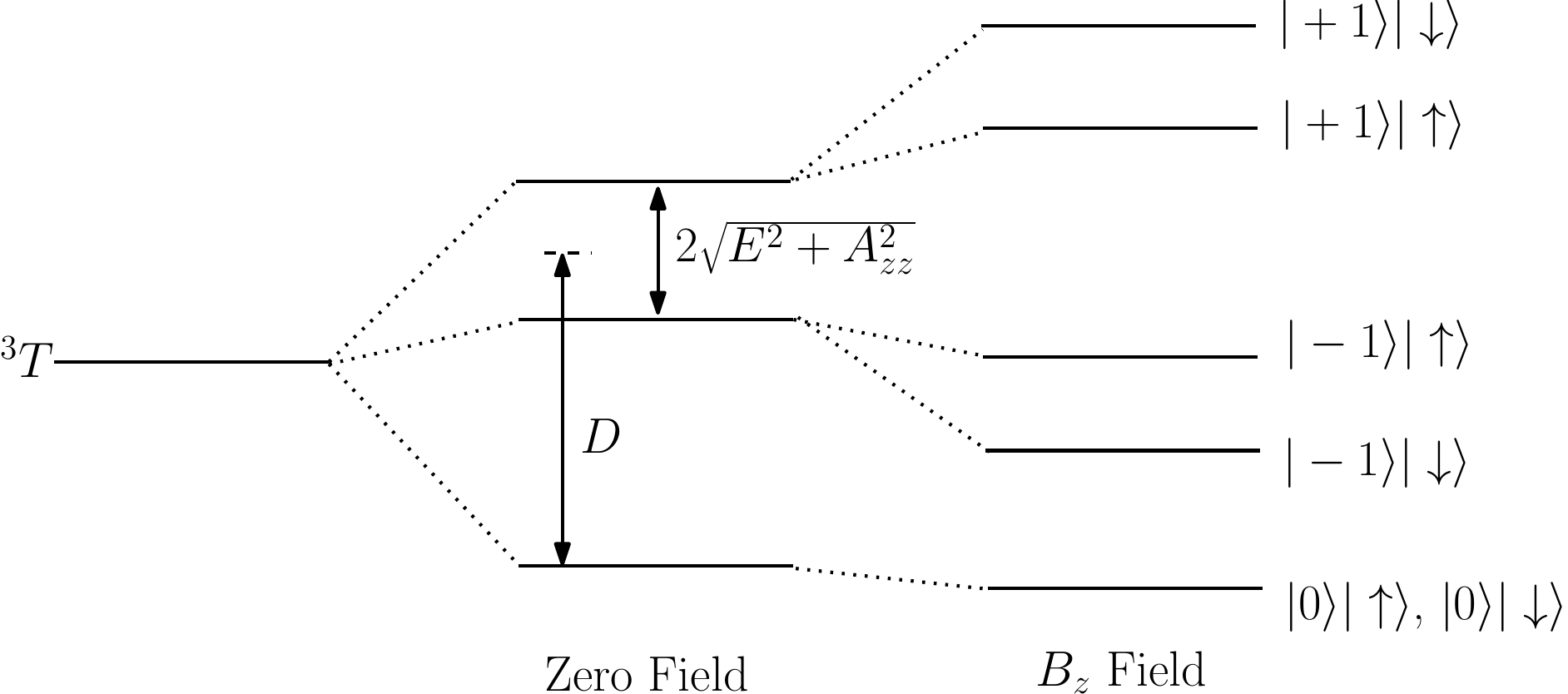}
  \caption{\textbf{Triplet level hyperfine structure.} The zero field energies are correct to first order in the
  hyperfine interaction. The approximate eigenstates in the presence of an aligned magnetic field fulfilling $B_z\gg E$
are expressed in terms of the electronic spin states ($|m_s\rangle$), where $m_s$ denotes electronic spin-projection,
and the nuclear spin states ($|\uparrow\rangle$, $|\downarrow\rangle$), where up/ down arrows denote $m_I=+\frac{1}{2}$/
$-\frac{1}{2}$ nuclear spin projections.}
  \label{fig:hyperfinestructure}
\end{figure}

\begin{figure}
  \centering
  \includegraphics[width=1\columnwidth]{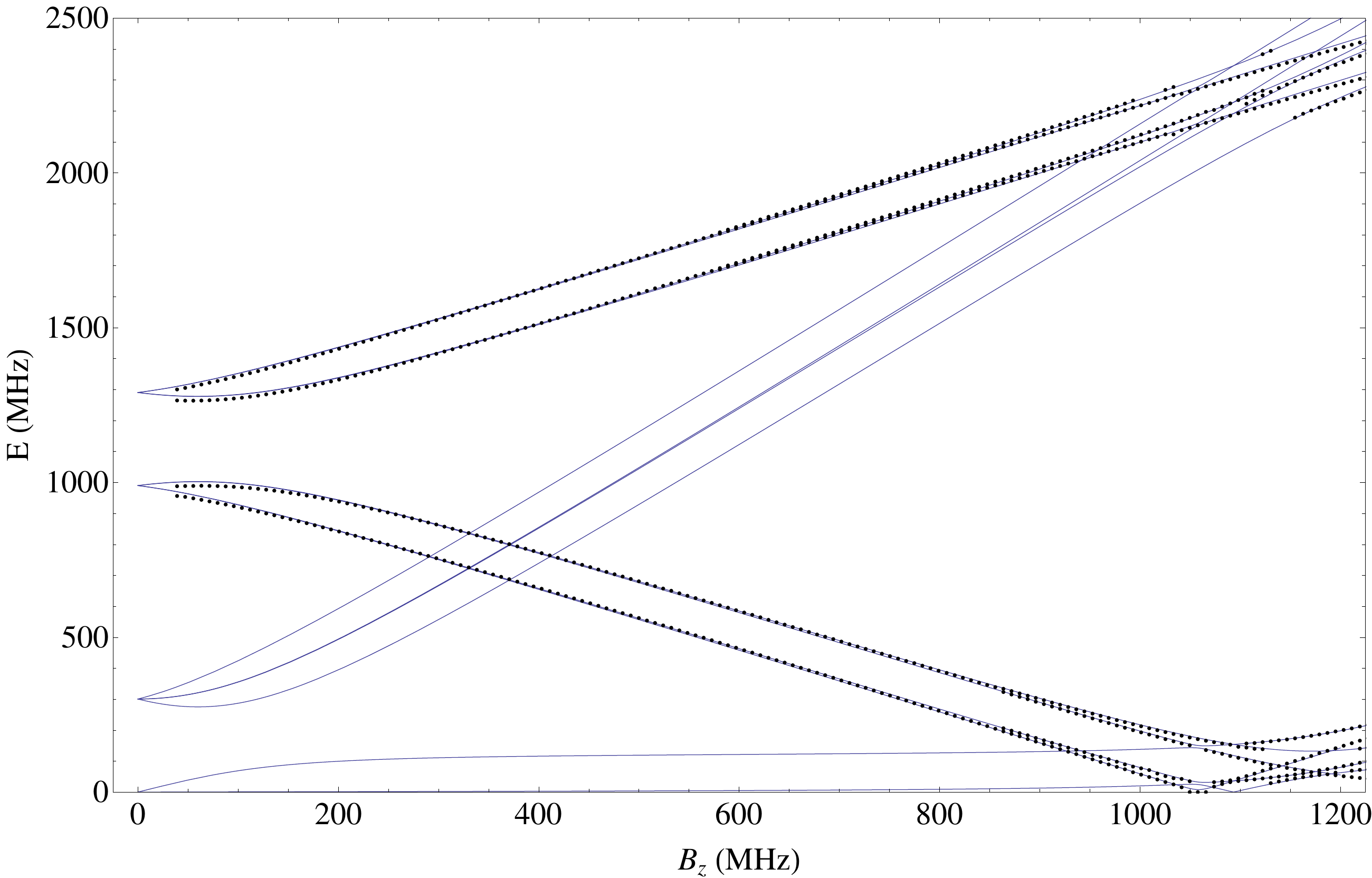}
  \caption{\textbf{Triplet hyperfine resonances as functions of $B_z$.} Solid lines denote the resonant frequencies
described by the spin-Hamiltonian model (\ref{eq:spinhamiltonian}, \ref{eq:hyperfineinteraction}). Points denote the
observed resonant frequencies.}
  \label{fig:hyperfinefit}
\end{figure}

Two of the ST1 defects studied exhibited hyperfine structure of the triplet level due to magnetic interactions with a
$^{13}C$ nucleus. One of them showed a particularly large hyperfine splitting greater than 100~MHz and was used
throughout this work (refer to figure~4 main text). Figure~\ref{fig:hyperfine_experiment} shows the hyperfine
splittings of this defect in the vicinity of a level-anti-crossing between the $|0\rangle$ and $|-1\rangle$ electronic
spin manifolds. These interactions are described by the addition of the following term to the triplet spin-Hamiltonian
(\ref{eq:spinhamiltonian}) \cite{Loubser1978,stoneham2001theory}
\begin{equation}
V_{hf} = \vec{S}\cdot\mathbf{A}\cdot\vec{I}
\label{eq:hyperfineinteraction}
\end{equation}
where $\vec{I}$ is the dimensionless $I=1/2$ nuclear spin operator and $\mathbf{A}$ is a $3\times3$ tensor to be
determined experimentally. With an appropriate choice of nuclear spin coordinate system $(XYZ)$, the tensor $\mathbf{A}$
may be diagonalized, simplifying the above to \cite{Loubser1978,stoneham2001theory}
\begin{equation}
V_{hf} = A_{xx}S_xI_X+A_{yy}S_yI_Y+A_{zz}S_zI_Z
\end{equation}
Note that the nuclear spin coordinate system $(XYZ)$ is in general different to the defect's coordinate system $(xyz)$.
Figure \ref{fig:hyperfinestructure} depicts the triplet hyperfine structure at zero field and in the presence of a
static magnetic field aligned along the defect's major axis ($B_z$) (nuclear Zeeman interaction ignored). At zero field,
the hyperfine structure is suppressed by the effects of the crystal field parameter $E$ (refer to figure~4 main text).
The observed hyperfine
resonances as functions of $B_z$ (data shown in figure~\ref{fig:hyperfine_experiment}), together with their least
squares best fit obtained using the spin-Hamiltonian model,
are depicted in figure \ref{fig:hyperfinefit}. The fitted values of the various spin-Hamiltonian parameters are
contained in table \ref{tab:hyperfineparameters}. Note that no difference could be detected between the hyperfine
parameters $A_{xx}$ and $A_{yy}$, such that they both can be equated to a single parameter $A_\perp=A_{xx}=A_{yy}$.

\begin{table}[htbp]
  \centering
    \begin{tabular*}{0.75\textwidth}{@{\extracolsep{\fill}}ccccccc}
    \hline
    $D$ (MHz) & $E$ (MHz) & $A_{zz}$ (MHz) & $A_\perp$ (MHz) \\
    \hline
    $±$1134.7(7) & $±$139(2) & -117(1) & -94(2) \\
    \hline
		\end{tabular*}%		
	\caption{\textbf{Triplet level hyperfine parameters.} The parameters are those of the single ST1 defect whose
hyperfine structure was studied in figure \ref{fig:hyperfinefit}. Note that $A_{xx}=A_{yy}=A_\perp$.}
  \label{tab:hyperfineparameters}%
\end{table}%

The hyperfine parameters can be used to interpret the character of the unpaired electronic spin density distribution at
the $^{13}C$ nucleus. The magnetic hyperfine interaction contains distinct Fermi contact $f$ and dipolar $d$ terms, such
that \cite{Loubser1978,He1993}
\begin{eqnarray}
A_{zz} & = & f+d \nonumber \\
A_\perp & = & f-2d
\end{eqnarray}
Using the fitted values of $A_{zz}$ and $A_\perp$ contained in table \ref{tab:hyperfineparameters}, the values
$f=-101(2)$ MHz and $d=-7.8(9)$ MHz can be determined.
The electronic orbitals $\psi$ that comprise the unpaired spin density associated with the $^{13}C$ atom can be well
described by linear combinations of $2s$ ($\phi_s$) and $2p$ ($\phi_p$) atomic orbitals \cite{Loubser1978,He1993}
\begin{equation}
\psi=c_s\phi_s+c_p\phi_p
\end{equation}
where $c_s$ and $c_p$ are coefficients that satisfy the normalization condition $1=|c_s|^2+|c_p|^2$. The Fermi contact
and dipolar terms may be expressed in terms of the orbital coefficients and the portion of unpaired spin density $\eta$
contributed by the atomic orbitals of the $^{13}C$ atom via \cite{Loubser1978,He1993}
\begin{eqnarray}
f = \frac{8\pi}{3}\frac{µ_0}{4\pi}g_eµ_Bg_nµ_n|c_s|^2\eta|\phi_s(0)|^2 \nonumber \\
d = \frac{2}{5}\frac{µ_0}{4\pi}g_eµ_Bg_nµ_n|c_p|^2\eta\langle\phi_p|\frac{1}{r^3}|\phi_p\rangle
\end{eqnarray}
where $g_n$ is the $^{13}C$ nuclear g-factor, $µ_n$ is the nuclear magneton, $\phi_s(0)$ is the value of the $2s$
orbital at the nucleus and $r$ is the distance of the electron spin from the nucleus. As per established practice in the
electron spin resonance of defects in diamond \cite{Loubser1978,stoneham2001theory, He1993, ayscough1967electron}, the
values of
$|\phi_s(0)|^2$ and $\langle\phi_p|\frac{1}{r}|\phi_p\rangle$ obtained from self-consistent field calculations of atomic
carbon can be used in conjunction with the observed values of $f$ and $d$ to estimate $|c_s|^2$, $|c_p|^2$ and $\eta$.
The estimated values are 0.27, 0.73 and 0.1, respectively.

The estimated values of $|c_s|^2$ and $|c_p|^2$ are very close to the values of a standard sp$^3$ bonding orbital of
diamond, being 0.25 and 0.75, respectively. Consequently, it appears that the spin density at the $^{13}C$ site of this
particular ST1 defect occupies a typical sp$^3$ orbital. Furthermore, only $\sim10\%$ of the spin density is associated
with that site, implying that the spin density is distributed over a number of atomic sites. This conclusion, when
combined with the observed $\sim1$ MHz ODMR linewidths and typical hyperfine interaction energies, implies that if the
defect possessed an intrinsic nuclear spin it would most likely have been detected. Hence, it may also be conservatively
concluded that the defect does not possess an intrinsic nuclear spin.

\section{Defect structure}

Summarising previous sections, the properties of the ST1 defect structure that we may conclude from our observations
are:
\begin{itemize}
    \item The optical transition occurs between energetically discrete electron orbitals within the diamond bandgap and
is accompanied by an appreciable lattice relaxation.
    \item The ground and optical excited electronic levels are spin singlets
    \item There exists a metastable spin triplet level between the singlet levels
    \item The triplet level's major axis is in the [110] direction.
    \item The defect has $C_{2v}$ or lower point group symmetry.
    \item The defect does not contain an intrinsic impurity that possesses a nuclear spin.
    \item The mean separation of the unpaired electrons of the triplet level is less than 4 angstroms.
    \item The unpaired spin density is distributed over a number of atomic sites.
    \item $10\%$ of the unpaired spin density is located at one particular carbon site and has the character of a sp$^3$
orbital.
\end{itemize}

\begin{figure}
  \centering
  \begin{subfigure}[b]{0.49\textwidth}
    \centering
    \includegraphics[width=0.75\columnwidth]{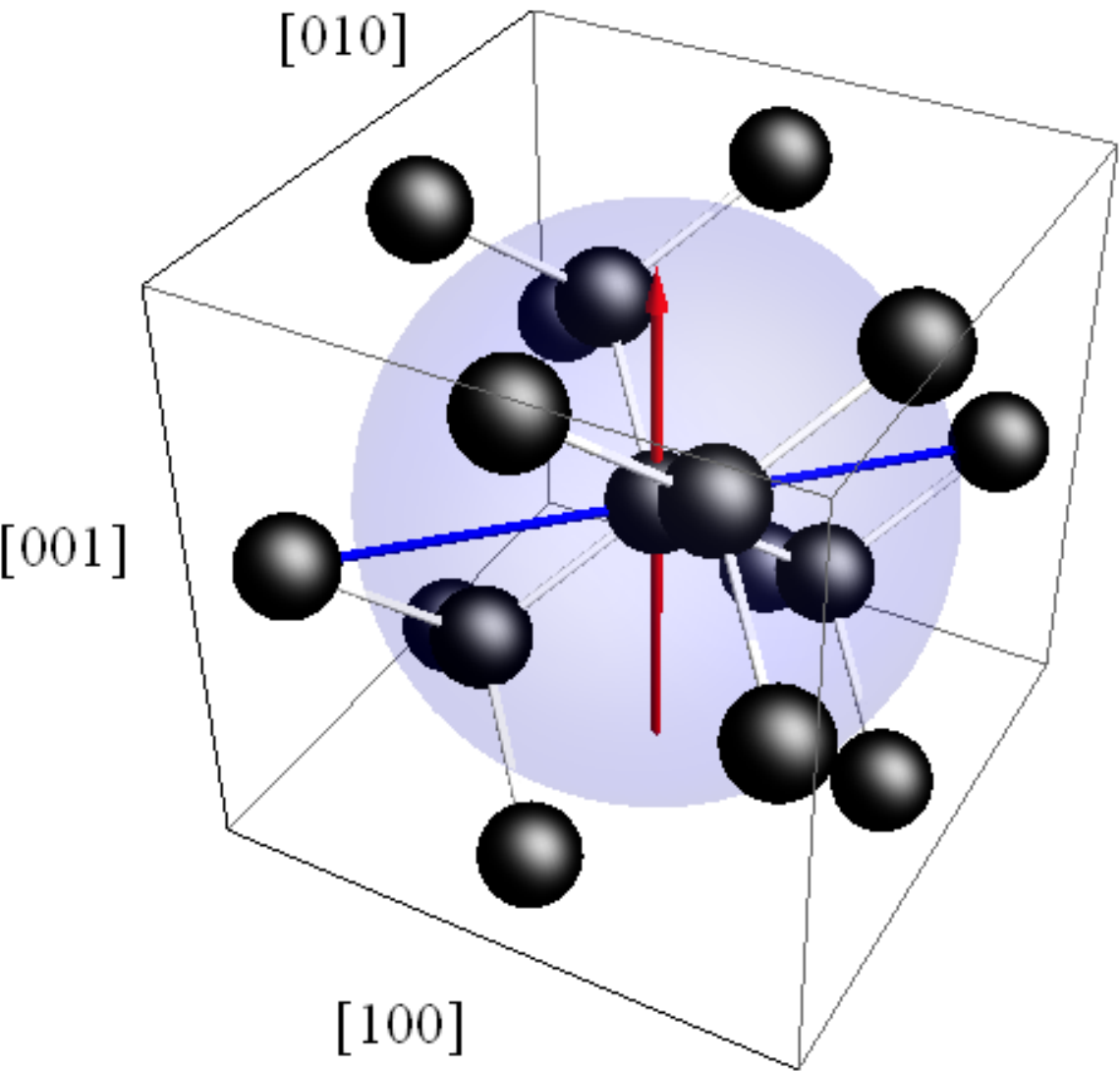}
  \end{subfigure}
  \begin{subfigure}[b]{0.49\textwidth}
    \centering
    \includegraphics[width=0.75\columnwidth]{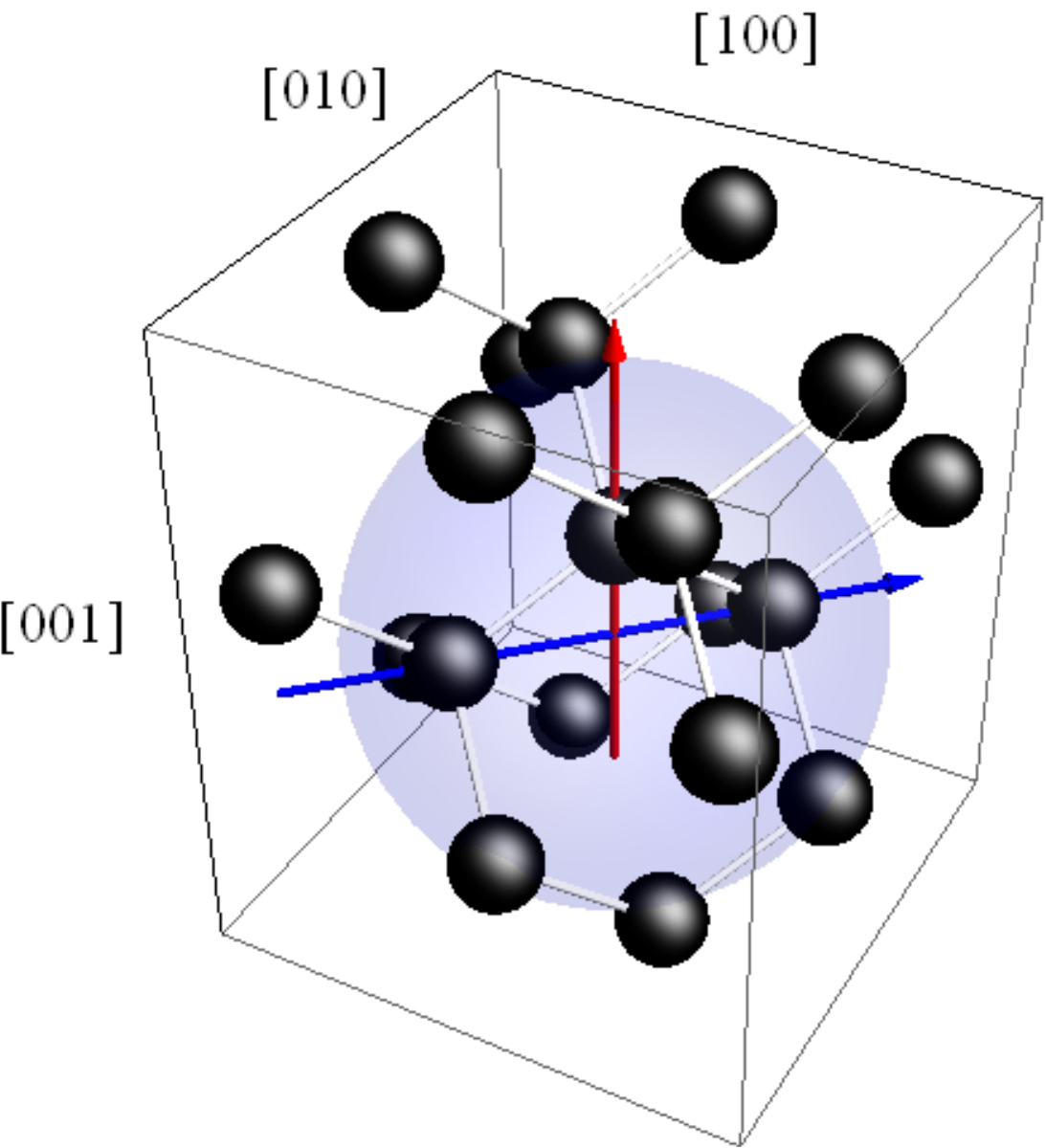}
  \end{subfigure}
  \caption{\textbf{Possible regions of the diamond lattice occupied by the ST1 defect.} The diagrams depict two portions
of the diamond crystal structure: (a) a portion centered on an atomic site and (b) a portion centered on the midpoint
between two atomic sites along the [110] direction. The transparent sphere in each diagram has a diameter of 4 angstroms
and indicates the region where the unpaired spin density of the triplet level is localized. The red arrow denotes the
$C_2$ symmetry axis of the [110] direction and the blue arrow is parallel to the [110] direction. The plane formed by
the $C_2$ symmetry axis and the [110] direction is the reflection plane of the diamond lattice corresponding to the
[110] direction. Black spheres represent atomic sites and white tubes denote sp$^3$ bonds.}
  \label{fig:defectstructure}
\end{figure}

Figure \ref{fig:defectstructure} depicts two portions of the diamond crystal structure: one portion is centered on an
atomic site and the other portion is centered on the midpoint between two atomic sites along the [110] direction. The
transparent sphere in each diagram has a diameter of 4 angstroms and thus indicates the region where the unpaired spin
density of the triplet level is localized, if the defect is atom centered or midpoint centered. Consistent with our
observations, it is clear that for both cases, there exist several atomic sites (up to 7) within the region where the
spin density is localized.

The conclusion that the defect does not contain an intrinsic impurity that possesses a nuclear spin restricts the set of
possible constituents of the defect. The restricted set contains substitutional and interstitial impurities without
nuclear spins, as well as the structural defects of diamond: the vacancy and self-interstitial. The defect is likely to
be a complex of these possible constituents, although given the relatively small region where the spin density is
localized, it is probable that it contains only up two or three. The simplest structure of the defect that is consistent
with our observations is a pair of constituents occupying atomic or interstitial sites along the [110] direction. A pair
of the same constituent will form a defect with the highest possible $C_{2v}$ symmetry if they are both occupying atomic
or interstitial sites. The pair of different constituents will form a defect with $C_{1h}$ symmetry.  More complicated
complexes can form defects with $C_2$ and $C_1$ symmetry.

Due to the involved fabrication process of the nano-wires, it is not possible to further restrict the set of possible
constituents of the defect. Systematic formation studies are required to gain more information about the constituents.
Given that oxygen plasma was used during the reactive ion etching process, it is capable of sp$^3$ bonding and the
abundant $^{16}O$ isotope does not have a nuclear spin, the presence of $^{16}O$ in the defect is certainly possible and
warrants further investigation. The significant strain that is likely to be present in the nano-wires potentially plays
a significant role in the formation energetics of the ST1 defect and may explain why the defect has not been
previously observed in bulk diamond.

\section{Nuclear spin polarization and readout}

\begin{figure}
\centering\includegraphics[width=1\columnwidth]{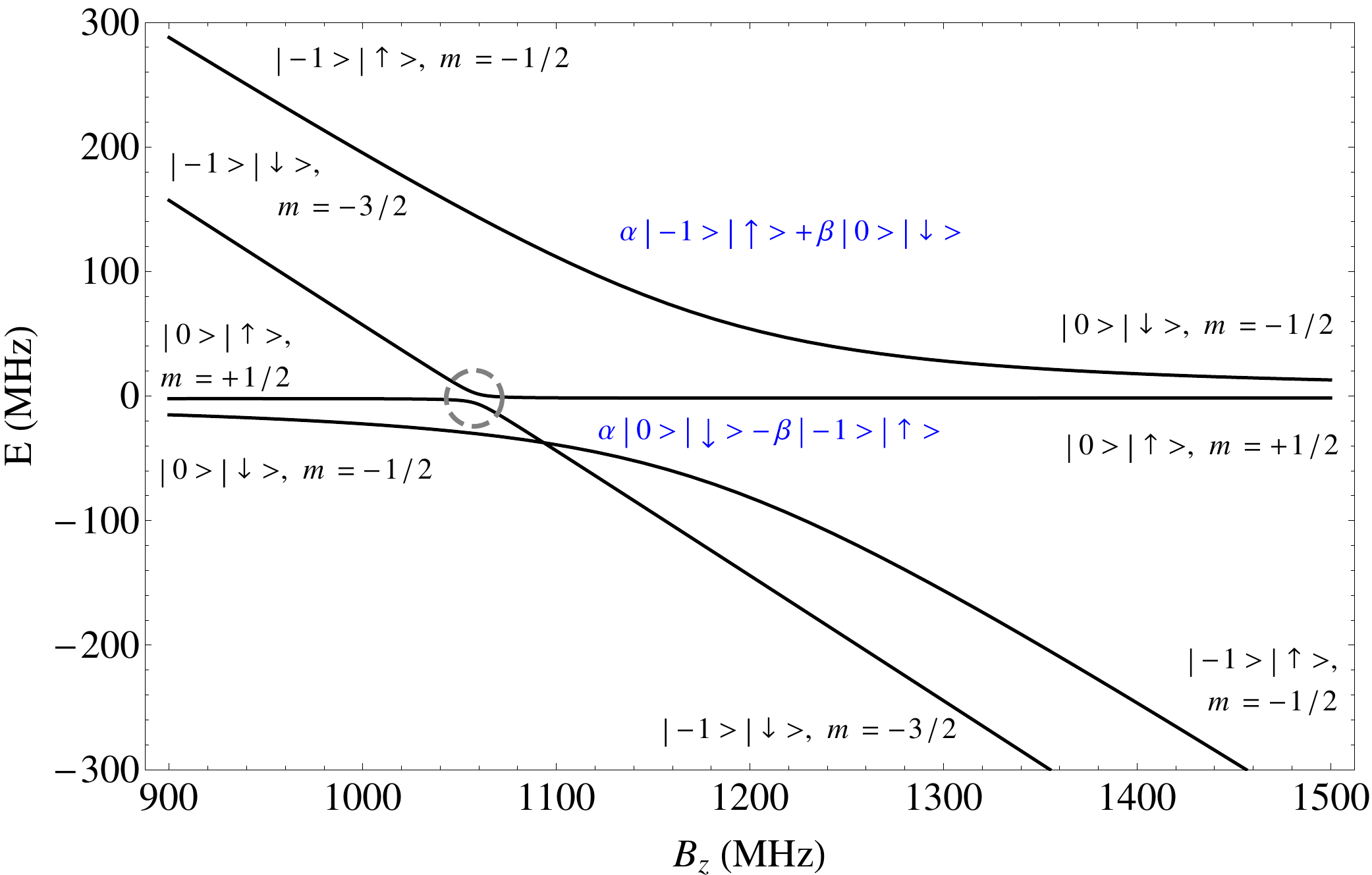}
\caption{\label{fig:lac}\textbf{The triplet level hyperfine level anti-crossing.} Solid lines denote the hyperfine energy levels in the region of the LAC. The approximate states and electro-nuclear spin projection $m$ of the levels before and after the LAC are depicted in black. The mixed states of the $m=-1/2$ levels at the LAC are depicted in blue. The dashed circle highlights the minor LAC between the $m=+1/2$ and $-3/2$ levels discussed in the text.}
\end{figure}

The hyperfine structure of the ST1 defect discussed in section \ref{sec:hyperfinestructure} exhibits a level anti-crossing (LAC) in the
region of a static magnetic field of strength $B_z\sim42$ mT that is aligned along the defect's major axis (refer to figure \ref{fig:lac}). Since $B_z$ and $D$ are
the dominate terms of the spin-Hamiltonian near the LAC, the electro-nuclear spin projection $m=m_s+m_I$ is an approximately good quantum and the levels
in figure \ref{fig:lac} have been labelled accordingly. In the region of the LAC, the transverse hyperfine terms $A_\perp$ mix levels of the same spin
projection $m$, such that the levels with $m=-\frac{1}{2}$ are mixed, but  the levels with $m=+\frac{1}{2}$ and $-\frac{3}{2}$ are
approximately unmixed. Strictly, the $m=+\frac{1}{2}$ and $-\frac{3}{2}$ levels do undergo a small LAC (as indicated in
figure \ref{fig:lac}) induced by the
relatively small $E$ terms. The $E$ term and anisotropic hyperfine coupling also result in small
admixtures of the $|+1,\pm 1/2\rangle$ hyperfine levels to the eigenstates near the LAC. We proceed by formulating a
basic model, where we neglect the $E$ term and assume isotropic hyperfine coupling. In this case, the mixed states
are composed from only two components $|-1,+1/2\rangle$ and $|0,-1/2\rangle$ (total spin $m=-\frac{1}{2}$) and they can
be quantified by the coefficients $\alpha$ and $\beta$ (with $|\alpha|^2+|\beta|^2=1$) as denoted in figure
\ref{fig:lac}.

\begin{figure}
\centering\includegraphics[]{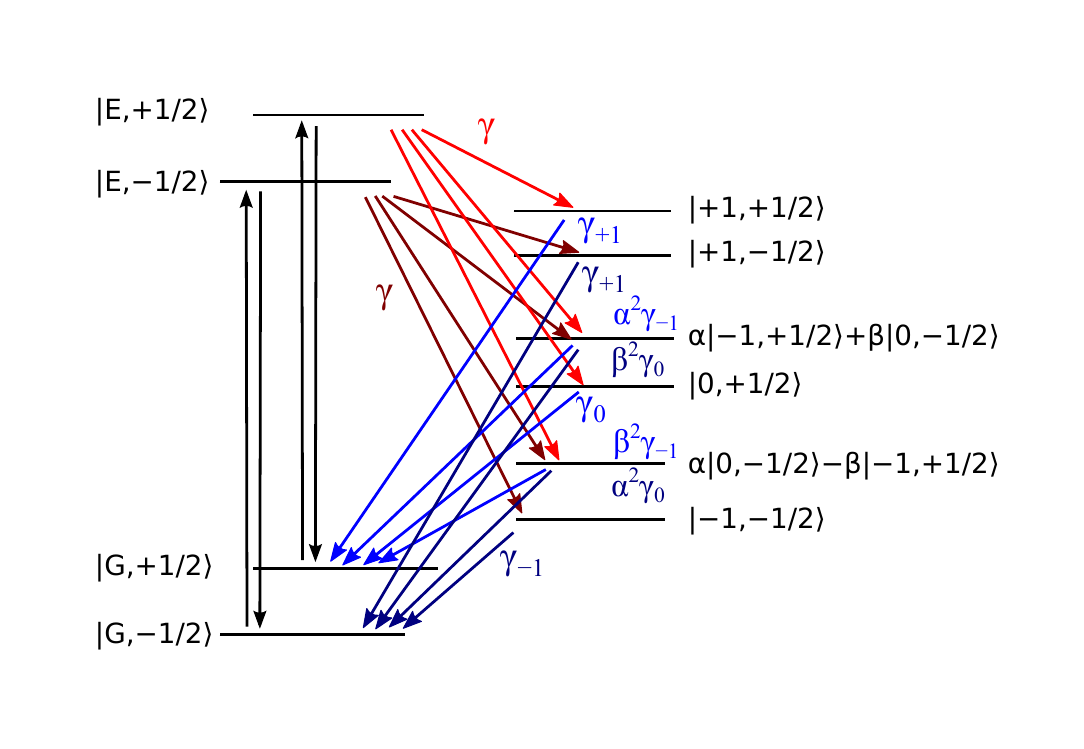}
\caption{\label{fig:onp}\textbf{10-level rate equation model for optical nuclear spin polarization and readout.}}
\end{figure}

\begin{figure}[!t]
\centering\includegraphics[width=0.7\columnwidth]{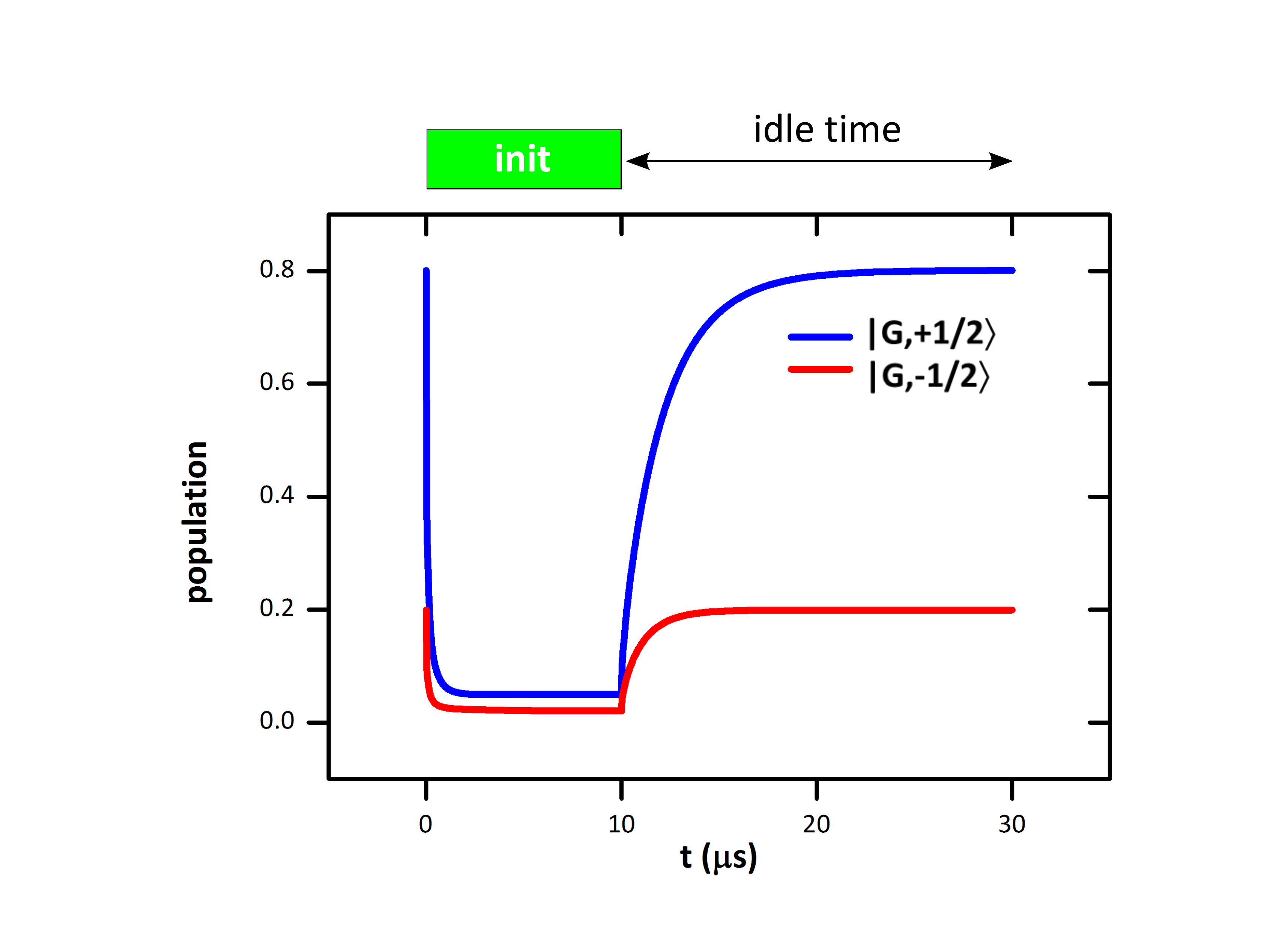}
\caption{\label{fig:transient}\textbf{Build-up of population in the ground state.} Each curve represents the time-evolution of two nuclear spin state populations during a pulse cycle. Note that 60~\% polarization is already achieved after the first cycle (not shown) and this plots shows a time-evolution after many cycles.}
\end{figure}

We now use this hamiltonian to construct a basic 10-level rate-equation model that describes nuclear spin polarization
by optical pumping and optical nuclear spin readout\cite{Stehlik1975,Colpa1977}.
We first neglect electronic spin relaxation and assume that the population rates of each of the triplet electronic spin
states are equal. This remaining transitions are summarized in figure~\ref{fig:onp}. To simplify the
notation, we denote the triplet population rate far away from the LAC by $\gamma$ and the decay rates out of the
electronic triplet states $\gamma_{0}$, $\gamma_{-1}$, $\gamma_{+1}$. The population and decay rates of the unmixed
states are
\begin{equation}
\begin{aligned}
  |E,\pm1/2\rangle\rightarrow|+1,\pm1/2\rangle&:\quad\gamma\\
  |E,+1/2\rangle\rightarrow|0,+1/2\rangle&:\quad\gamma\\
  |E,-1/2\rangle\rightarrow|-1,-1/2\rangle&:\quad\gamma\\
  |+1\pm1/2\rangle\rightarrow|G,\pm1/2\rangle&:\quad\gamma_{+1}\\
  |0,+1/2\rangle\rightarrow|G,+1/2\rangle&:\quad\gamma_{0}\\
  |-1,-1/2\rangle\rightarrow|G,-1/2\rangle&:\quad\gamma_{-1}\\
\end{aligned}
\end{equation}
The transitions rates for the mixed states $|I\rangle=\alpha|-1,+1/2\rangle+\beta|0,-1/2\rangle$ and
$|II\rangle=\alpha|0,-1/2\rangle-\beta|-1,+1/2\rangle$ are easily found by considering the matrix elements of the
transition operator $D$. For example the decay rate from $|I\rangle$ to ground state $+1/2$ is found
as
\begin{equation}
\begin{aligned}
  \gamma_{|I\rangle|G,+1/2\rangle}&=|\langle G,+1/2|D(\alpha|-1,+1/2\rangle+\beta|0,-1/2\rangle)|^2\\
  &=|\alpha\langle G,+1/2|D|-1,+1/2\rangle+\beta\langle G,+1/2|D|0,-1/2\rangle|^2\\
  &=|\alpha\langle G,+1/2|D|-1,+1/2\rangle|^2=|\alpha|^2\gamma_{-1}
\end{aligned}
\end{equation}
The population and decay rates of the mixed states are thus
\begin{equation}
\begin{aligned}
  |E,+1/2\rangle\rightarrow\alpha|-1,+1/2\rangle+\beta|0,-1/2\rangle&:\quad|\alpha|^2\gamma\\
  |E,-1/2\rangle\rightarrow\alpha|-1,+1/2\rangle+\beta|0,-1/2\rangle&:\quad|\beta|^2\gamma\\
  |E,+1/2\rangle\rightarrow\alpha|0,-1/2\rangle-\beta|-1,+1/2\rangle&:\quad|\beta|^2\gamma\\
  |E,-1/2\rangle\rightarrow\alpha|0,-1/2\rangle-\beta|-1,+1/2\rangle&:\quad|\alpha|^2\gamma\\
  \alpha|-1,+1/2\rangle+\beta|0,-1/2\rangle\rightarrow|G,+1/2\rangle&:\quad|\alpha|^2\gamma_{-1}\\
  \alpha|-1,+1/2\rangle+\beta|0,-1/2\rangle\rightarrow|G,-1/2\rangle&:\quad|\beta|^2\gamma_0\\
  \alpha|0,-1/2\rangle-\beta|-1,+1/2\rangle\rightarrow|G,+1/2\rangle&:\quad|\beta|^2\gamma_{-1}\\
  \alpha|0,-1/2\rangle-\beta|-1,+1/2\rangle\rightarrow|G,-1/2\rangle&:\quad|\alpha|^2\gamma_0\\
\end{aligned}
\end{equation}
Spin polarization proceeds in two steps. First, we perform optical pumping until steady state is reached. Thereafter
the system is allowed to decay to the ground state. Before we evaluate the numerical solution to the rate-equations,
we provide a qualitative argument and consider the case at the center of the LAC, where $\alpha=\beta=2^{-1/2}$. Under
this condition, the total population rates of all triplet hyperfine states are equal. Thus, the steady state population
is determined by the decay rates. Specifically, population will be trapped predominantly in the longest lived hyperfine
level. The lifetimes $\tau_i=1/\gamma_i$ of the triplet states are
\begin{equation}
\begin{aligned}\label{eq:hyperfine_lifetimes}
 |+1,\pm1/2\rangle&:\quad\gamma_{+1}^{-1}\approx 200 ns\\
 \alpha|-1,+1/2\rangle+\beta|0,-1/2\rangle&:\quad(1/2\gamma_{0}+1/2\gamma_{-1})^{-1}\approx 1400 ns\\
 |0,+1/2\rangle&:\quad\gamma_{0}^{-1}\approx 2500 ns\\
 \alpha|0,-1/2\rangle+\beta|-1,+1/2\rangle&:\quad(1/2\gamma_{0}+1/2\gamma_{-1})^{-1}\approx 1400 ns\\
 |-1,-1/2\rangle&:\quad\gamma_{-1}^{-1}\approx 1000 ns\\
\end{aligned}
\end{equation}
The $|+1\pm 1/2\rangle$ states do not contribute imbalance of the decay rates for the $+1/2$ and $-1/2$ nuclear spin
manifolds and can be ignored. The role of the mixed states is twofold. First, they provide nuclear spin flip
transitions, and second, due to their larger decay rates to the $+1/2$ ground state (see above), they result in an
overall increased rate from the $-1/2$ to the $+1/2$ manifold. The unmixed states provide nuclear spin polarization by
their different lifetimes. Specifically, population will accumulate predominantly in the long lived state
$|0,+1/2\rangle$. Both mechanisms act in the same direction, such that, in the steady state under strong optical
pumping, the largest fraction of the population will be in $|0,+1/2\rangle$, a relatively large population will also be
in the mixed states, minor population will be in the $|-1,+1/2\rangle$ state and negligable population in the $|+1,\pm
1/2\rangle$ states. During a subsequent decay (after the pump laser is turned off), all population from the
$|0,+1/2\rangle$ will decay to the $+1/2$ ground state and the mixed states will preferentially decay into the $+1/2$
ground state. This will further increase the nuclear spin polarization. This process is described in terms of the change in the two ground state populations during a pulse sequence in figure~\ref{fig:transient}. A full numerical analysis predicts a nuclear spin polarization $(n_{+1/2}-n_{-1/2})/(n_{+1/2}+n_{-1/2})\approx 60\%$.

We now consider nuclear spin readout. Starting from ground state $+1/2$ or ground state $-1/2$, we apply optical
pumping until steady state is reached and evaluate the integrated fluorescence, that is given by the excited state
populations (multiplied by the optical decay rate). The nuclear spin states will have different probability to be
trapped in the long lived shelving state $|0,+1/2\rangle$ and thus have a different overall fluorescence intensity.
The mixed states do not contribute to readout contrast, since their population rates are equal and as soon as they are
populated, the nuclear spin information is lost. Likewise, the $|+1,\pm 1/2\rangle$ states do not contribute to
readout contrast and the only relevant states are the two unmixed states. Appreciating the lifetimes evaluated
in Eq.(\ref{eq:hyperfine_lifetimes}), we expect that the nuclear spin $|G,+1/2\rangle$ will correspond to low
fluorescence. A full numerical analysis predicts a fluorescence contrast between $|G,+1/2\rangle$ and $|G,-1/2\rangle$
of about 7\%. The numerical analysis also predicts that the time to reach steady state is about $6-8 \mu$s, as
observed in experiment.

We now combine the results for nuclear spin polarization and readout contrast. The nuclear spin signal will be the
product of both efficiencies. From this we expect a signal contrast of about 4\%. This slightly overestimates the
experimentally observed contrast of about 1.5 \%. This likely reflects the approximations made in this basic
model. Next, we can consider the sign of the signal contrast contrast. Experimentally, we observe a maximum of the
nuclear Rabi oscillations at zero microwave pulse length, indicating that the nuclear spin state that we polarize in
corresponds to bright fluorescence. By contrast, the basic model suggests the opposite case, where the polarized
state ($|0,+1/2\rangle$) corresponds to low fluorescence. This is likely also a result of the simplifications made in
above basic model.

Several aspects are likely to be relevant in a full description of the nuclear spin polarization- and readout contrast.
One aspect is the small admixture from the $|+1\pm1/2\rangle$ states into the eigenstates at the LAC. Due to the
comparably fast decay rate of the these states, the relatively small admixture might become relevant. A second
aspect are possible small differences among the population rates. While these are not important for the electron spin
polarization, here they could become relevant. After designing a precise measurement for the population rates far away
from the LAC, these could be included in a complete rate model. Lastly, electron-spin relaxation might be considered.
Electron-spin relaxation would result in additional transition rates among the triplet hyperfine levels.

\bibliographystyle{unsrt}
\bibliography{library}

\end{document}